\documentclass{aastex62}

\usepackage{amsmath}

\usepackage{graphicx}	

\usepackage{xcolor}

\renewcommand{\vec}{\textbf}

\received{January 1, 2018}
\revised{January 7, 2018}
\accepted{\today}
\submitjournal{ApJS}

\shorttitle{Arepo public release}
\shortauthors{R. Weinberger, V. Springel \& R. Pakmor}

\begin{document}

\title{The {\sc Arepo} public code release}

\correspondingauthor{rainer.weinberger@cfa.harvard.edu}
\email{}

\author{Rainer Weinberger}
\affil{Center for Astrophysics $\mid$ Harvard \& Smithsonian, 
60 Garden Street, Cambridge, MA 02138, USA}

\author{Volker Springel}
\author{R\"udiger Pakmor}
\affiliation{Max-Planck Institute for Astrophysics, 
Karl-Schwarzschild-Str. 1, 
D-85741 Garching, Germany
}

\collaboration{}



\begin{abstract}

  We introduce the public version of the cosmological
  magnetohydrodynamical moving-mesh simulation code
  \textsc{Arepo}. This version contains a finite-volume
  magnetohydrodynamics algorithm on an unstructured, dynamic Voronoi
  tessellation coupled to a tree-particle-mesh algorithm for the
  Poisson equation either on a Newtonian or cosmologically expanding
  spacetime. Time integration is performed adopting local time step constraints
  for each cell individually, solving the fluxes only across active
  interfaces, and calculating gravitational forces only between active
  particles, using an operator-splitting approach. This allows
  simulations with high dynamic range to be performed
  efficiently. \textsc{Arepo} is a massively distributed-memory
  parallel code, using the message passing interface (MPI)
  communication standard and employing a dynamical workload and
  memory balancing scheme to allow for optimal use of multi-node parallel
  computers. The employed parallelization algorithms of \textsc{Arepo}
  are deterministic and produce binary-identical results when rerun
  on the same machine and with the same number of MPI ranks. A simple
  primordial cooling and star formation model is included as
  an example of sub-resolution models commonly used in simulations of
  galaxy formation. \textsc{Arepo} also contains a suite of
  computationally inexpensive test problems, ranging from idealized
  tests for automated code verification to scaled-down versions of
  cosmological galaxy formation simulations, and is extensively
  documented in order to assist adoption of the code by new scientific
  users.
\end{abstract}



\section{Introduction} \label{sec:intro}

%

Computer simulations have become an essential part of modern
astrophysical research \citep[e.g.][]{Naab+2016}. They allow for the
numerical determination of the time evolution of initial-value
problems, governed by (almost) arbitrarily complex partial
differential equations. Over the past decades, computing power
available for this approach has grown enormously, accelerating
progress in the field by allowing ever more complex systems to be
simulated with increasing levels of physical fidelity. However,
continued rapid progress is only possible if advancements in computer
hardware go hand in hand with the development of more sophisticated
simulation software, as well as with improvements of the employed
numerical methods \citep[][for reviews of some of the
techniques]{Springel2010b, Lehner+2014,Teyssier2015}. The former
involves efficient parallelization to achieve workload and memory
balance on distributed-memory machines with a minimum amount of
communication, something that can be quite nontrivial to achieve for
tightly coupled problems. The latter involves the use of more accurate
and flexible numerical schemes, and in particular the ability to focus
the computational efforts on regions of particular interest. 

Because the simulation codes themselves become increasingly complex, a third
challenge emerges in the form of a need to better organize the
improvement of software capabilities in (potentially large)
collaborative teams \citep{PortegiesZwart2018}. This calls for
detailed documentation, modularity and portability of astrophysical
simulation software, as well as for a continuous verification of
individual parts contributed by scientific users with often very
diverse technical backgrounds and development styles.

In this paper, we describe the public version of the \textsc{Arepo}
code,\footnote{homepage: www.arepo-code.org, repository: gitlab.mpcdf.mpg.de/vrs/arepo}  
a multipurpose gravitational and finite-volume magnetohydrodynamical (MHD) 
code for cosmic structure formation and more general astrophysical problems. 
\textsc{Arepo} was originally presented in \citet{Springel2010}, and has 
subsequently been applied to a wide range of astrophysical problems. 
These include cosmological simulations of galaxy formation 
{\citep[see][for a recent review]{Vogelsberger+2019}} in large volumes 
such as Illustris \citep{Genel+2014, Vogelsberger+2014} and IllustrisTNG
\citep{Marinacci+2018, Naiman+2018, Nelson+2018, Nelson+2019, 
  Pillepich+2018, Pillepich+2019, Springel+2018}, cosmological zoom
simulations of galaxy formation such as Auriga \citep{Grand+2017},
isolated galaxies \citep{Jacob+2018}, spiral structure in galaxies
\citep{Smith+2014}, stratified box simulations modeling a part of a
disk galaxy \citep{Simpson+2016}, wind-tunnel-like setups
\citep{Sparre+2019}, turbulent boxes \citep{Bauer+2012, Mocz+2017},
SNe Ia \citep{Pakmor+2013b}, binary stars in a common envelope 
phase \citep{Ohlmann+2016}, tidal disruption events \citep{Goicovic+2019},
protoplanetary and accretion disks \citep{Munoz+2014, Fiacconi+2018}
and astrophysical jets \citep{Bourne+2017, Weinberger+2017}.

These scientific applications have become possible thanks to a large
number of people contributing to expanding and further refining
\textsc{Arepo} since its initial description by \citet{Springel2010},
and by making the code available upon request to a sizable number of
people outside the group of direct scientific collaborators of the
original \textsc{Arepo} developers. However, the code had not been
made publicly available to the full astronomical community thus far,
an undesirable situation that we aim to address here. We share the
view that making simulation codes publicly available constitutes good
scientific practice in computational astrophysics, as it supports
transparency and reproducibility, helps to identify bugs more quickly,
and ultimately accelerates progress in the field by allowing
scientists to build up more easily on previous work. Other codes in
cosmological structure formation have led the way in this regard, such
as \textsc{Gadget} \citep{Springel+2001}, \textsc{Ramses}
\citep{Teyssier+2002}, \textsc{Athena} \citep{Stone+2008},
\textsc{Enzo} \citep{Bryan+2014}, \textsc{Changa} \citep{Menon+2015},
and \textsc{Phantom} \citep{Price2018}, to name just a few.

The public release version of \textsc{Arepo} presented here
corresponds to the state of the master development branch of the code
as of November 2017, albeit with somewhat reduced functionality with
respect to special modules and features. This was done in order to
provide a significantly simplified and completely documented code,
both at a user and a developer level, in order to enable scientists to
use and extend the code without direct support by the authors. This
paper serves as an overview of the available functionality, and it
summarizes a number of the updates done over the years since the
initial exposition of the code in \citet{Springel2010}. It can also be
read as an introduction of \textsc{Arepo} to new users. To facilitate
the latter, we connect used variables in equations with the
corresponding code parameter names, where appropriate. In the interest
of conciseness, we refer interested readers wherever applicable to the
original development papers for in-depth discussions on why certain
implementations were chosen over alternative methods.

This paper is structured as follows. In \S2 the basic equations are
presented. We discuss the solvers for gravitational interactions and
MHD in \S3 and \S4, respectively. In \S5 we discuss the creation and
dynamics of the computational mesh on which the MHD equations are
discretized. \S6 describes the implementation of additional source and
sink terms such as radiative cooling as well as the coupling to
sub-resolution models using star-formation as an example. In \S7 we
present the time integration, and in \S8, we present the dynamic workload and
memory balancing scheme in \textsc{Arepo}. \S9 covers other aspects of
the code, such as the on-the-fly subhalo identification and initial
conditions creation as well as important implementation aspects of
the code such as input and output, memory management, and the
use of external libraries.  Finally, we discuss a number of examples
and test cases in \S10, present the code development and support strategy
in \S11, and give a summary in \S12.

\section{Equations}
\label{sec:equations}

\textsc{Arepo} solves the equations of a collisionless particle
component and of hydrodynamics on a uniformly expanding, flat
Friedmann-Lema\^itre-Robertson-Walker spacetime. The space expansion is described by a single quantity, the scale factor $a(t)$, defined by the line element \citep[see, e.g.][]{Mo+2010}
\begin{align}
{\rm d}s^2 = c^2 {\rm d}t^2 - a^2(t) \left[ {\rm d}r^2 + r^2
  \left({\rm d}\vartheta^2 + \sin^2\vartheta \, {\rm d} \varphi^2 \right) \right] .
\end{align}

Co-moving quantities are defined as follows:
\begin{align}
  \vec{x}_c &= \vec{x}/a, \\
  \vec{v}_c &=\vec{v} - \dot{a} \, \vec{x}_c ,\\
  \rho_c &= a^{3} \rho ,\\
  p_c &= a^{3} p ,\\
  \Phi_c &= a \Phi + \ddot{a} a^2 \vec{x}_c^2/2, \\
  \vec{B}_c &= a^{2} \vec{B},
\end{align}
where $\vec{x}$ and $\vec{v}$ are the ``proper" position and velocity
vectors, respectively, $\rho$ is the density, $p$ is the pressure, $\Phi$ is the
gravitational potential, and $\vec{B}$ is the magnetic field strength. The
variables with a subscript $c$ denote the respective ``co-moving''
quantities, which are introduced to absorb the evolution of these
quantities due to space expansion in full or in part, and thus to
simplify the corresponding equations.  To transform equations from
proper to co-moving variables, it is important to observe that not
only the coordinates but also the time and spatial derivatives change
according to
\begin{align}
\frac{\partial f}{\partial t} = \left. \frac{\partial f}{\partial t}\right|_{\vec{x}_c} &= \left.\frac{\partial  f}{\partial t}\right|_{\vec{x}} + \frac{\dot{a}}{a} \vec{x}_c \cdot  \nabla_c f , \\
\nabla_c f &= a \nabla f .
\end{align}
Note that a static spacetime can be easily recovered from the
cosmological form of the equations by setting $a=1$ and
$\dot{a}=\ddot{a}=0$. We therefore present the equations with co-moving
variables in the following, dropping the subscript $c$ for simplicity.

\subsection{Cosmological background evolution}

The evolution of the scale factor $a$ is given by the Friedmann equation, assuming a  $\Lambda$ cold dark matter cosmology:
\begin{align}
  H = H_0 \, [\Omega_0 a^{-3} + (1 - \Omega_0 - \Omega_\Lambda)
  a^{-2} + \Omega_\Lambda]^{1/2} ,
\end{align}
with $H=\dot{a} a^{-1}$ and where $\Omega_0$ is the matter density in
the universe relative to the critical density, and $\Omega_\Lambda$
represents the corresponding density parameter for the cosmological
constant. Note that the contribution of radiation to the cosmic
expansion history can be neglected for the redshifts of interest.

\subsection{Gravitational potential}

The gravitational potential $\Phi$ in an expanding spacetime can be obtained from the Poisson equation in proper space, replacing the variables with the co-moving ones, and using the Friedmann equation to connect second derivatives of the scale factor to the mean density,
 \begin{align}
    \nabla^2 \Phi = 4 \pi G  \left( \rho_\text{total} -
   \rho_\text{mean} \right) ,
    \label{eq:Poisson}
  \end{align}
  with $G$ being the gravitational constant, and $\rho_\text{total}$ and $\rho_\text{mean}$ are the total and mean density, respectively.

\subsection{Collisionless fluid}

The collisionless component (i.e.~dark matter and stars in galaxies)
can be described by the Vlasov equation
\begin{align}
  \frac{{\rm d} f}{{\rm d}t} &= \frac{\partial f}{\partial t} + \frac{\partial f}{\partial \vec{x}} \frac{\partial \vec{x}}{\partial t} + \frac{\partial f}{\partial \vec{v} } \frac{\partial \vec{v}}{\partial t} \\
   &= \frac{\partial f}{\partial t} + \frac{\partial f}{\partial \vec{x}}\, \frac{\vec{v}}{a} - \frac{\partial f}{\partial \vec{v} } \left( \frac{\nabla \Phi}{a^2} + \frac{\dot{a}}{a} \vec{v} \right)  = 0.
   \label{eq:Vlasov}
\end{align}
 Applying the method of characteristics yields simple equations of motion,
  \begin{align}
    \dot{\vec{x}} &= \vec{v}/a , \label{eom1} \\
    \dot{\vec{v}} &=  - \frac{\nabla \Phi }{a^2} - \frac{\dot{a}}{a} \vec{v} ,\label{eom2}
  \end{align}
which can be used to integrate the trajectories of discrete particles
that sample the initial phase-space distribution function.

\subsection{Ideal Magnetohydrodynamics}

The  equations of ideal magnetohydrodynamics (MHD) can be written as \citep{Pakmor+2013}
\begin{align}
  \frac{\partial \rho}{\partial t} &+ \frac{1}{a} \nabla \cdot \left( \rho \vec{v} \right) = 0, \label{eq:euler:mass}\\
  \frac{\partial \rho \vec{w} }{\partial t} &+ \nabla \cdot \left(
                                              \rho \vec{v} \vec{v}^{T}
                                              + \vec{I} p_\text{tot} -
                                              \frac{\vec{B}
                                              \vec{B}^{T}}{a} \right)
                                              = - \frac{\rho}{a} \nabla
                                              \Phi \label{eq:euler:momentum}
  ,\\
  \frac{\partial \mathcal{E} }{\partial t} &+ a \nabla \cdot \left[ \vec{v} \left( E + p_\text{tot} \right) - \frac{1}{a} \vec{B} \left( \vec{v} \cdot \vec{B} \right) \right] = \frac{\dot{a}}{2} \vec{B}^2 \label{eq:euler:energy} - \rho \left( \vec{v}\cdot \nabla \Phi\right) + a^2 \left(\mathcal{H} - \Lambda \right), \\
  \frac{\partial \vec{B}}{\partial t} &+ \frac{1}{a} \nabla \cdot \left( \vec{B} \vec{v}^{T} - \vec{v} \vec{B}^{T} \right) = 0. \label{eq:euler:induction}
\end{align}
Note that the momentum and energy equations take a slightly different
form compared to, e.g., those of \citet{Bryan+2014}. We prefer this formulation,
however, as the use of time derivatives of the auxiliary variables
\begin{align}
  \vec{w} &= a \vec{v} ,\\
  \mathcal{E} &= a^2 E,
\end{align}
 reduces the number of MHD source terms to a single magnetic field term in the energy equation.
The total energy and pressure is calculated as
\begin{align}
  E &=  \rho u_\text{th} + \frac{1}{2} \rho \vec{v}^2 + \frac{ \vec{B}^2 }{2 a} ,\\
  p_\text{tot} &= \left(\gamma-1\right) \rho u_\text{th} + \frac{\vec{B}^2}{2 a}.
\end{align}
Note that possible external heating and cooling, described through
$a^2 \left(\mathcal{H} - \Lambda \right)$,
appear as source terms in the energy equation~(\ref{eq:euler:energy}).

\section{Modeling gravitational interactions via $N$-body dynamics}
\label{sec:gravity}

The Vlasov equation (\ref{eq:Vlasov}) describes the time evolution of
the phase-space distribution function of a collisionless
fluid. However, the high dimensionality of this equation makes direct
discretization on a grid a computationally challenging task
\citep{Yoshikawa+2013}. Therefore, it is useful to sample the phase-space 
density $f$ via discrete particles and solve the equations of
motion of these simulation particles (equations \ref{eom1} and
\ref{eom2}). The key computational bottleneck is the computation of
the acceleration $-\nabla \Phi$ of each individual particle because
the potential $\Phi$ is generated from all other particles. This leads
to, in its pure form, a computational effort scaling with $N^2$, where
$N$ is the number of simulation particles. To avoid this unfavorable
scaling behavior with particle number, \textsc{Arepo} uses two
different techniques. The first one is an oct-tree algorithm, which
groups distant particles and calculates their collective contribution
to the overall force, while the second method is a grid-based
approach, where the gravitational force is calculated on a Cartesian
grid via Fourier methods and then interpolated to each particle
position. Each of these techniques has its advantages and
disadvantages. Therefore, \textsc{Arepo} offers to use a combination
of both techniques, combining the efficiency and implicit periodicity
of a particle-mesh algorithm with the ability to cover a large dynamic
range, which is an inherent strength of the tree algorithms.

\subsection{Hierarchical multipole expansion with an oct-tree algorithm}

\begin{figure}
\centering
  \includegraphics[width=1.0\textwidth]{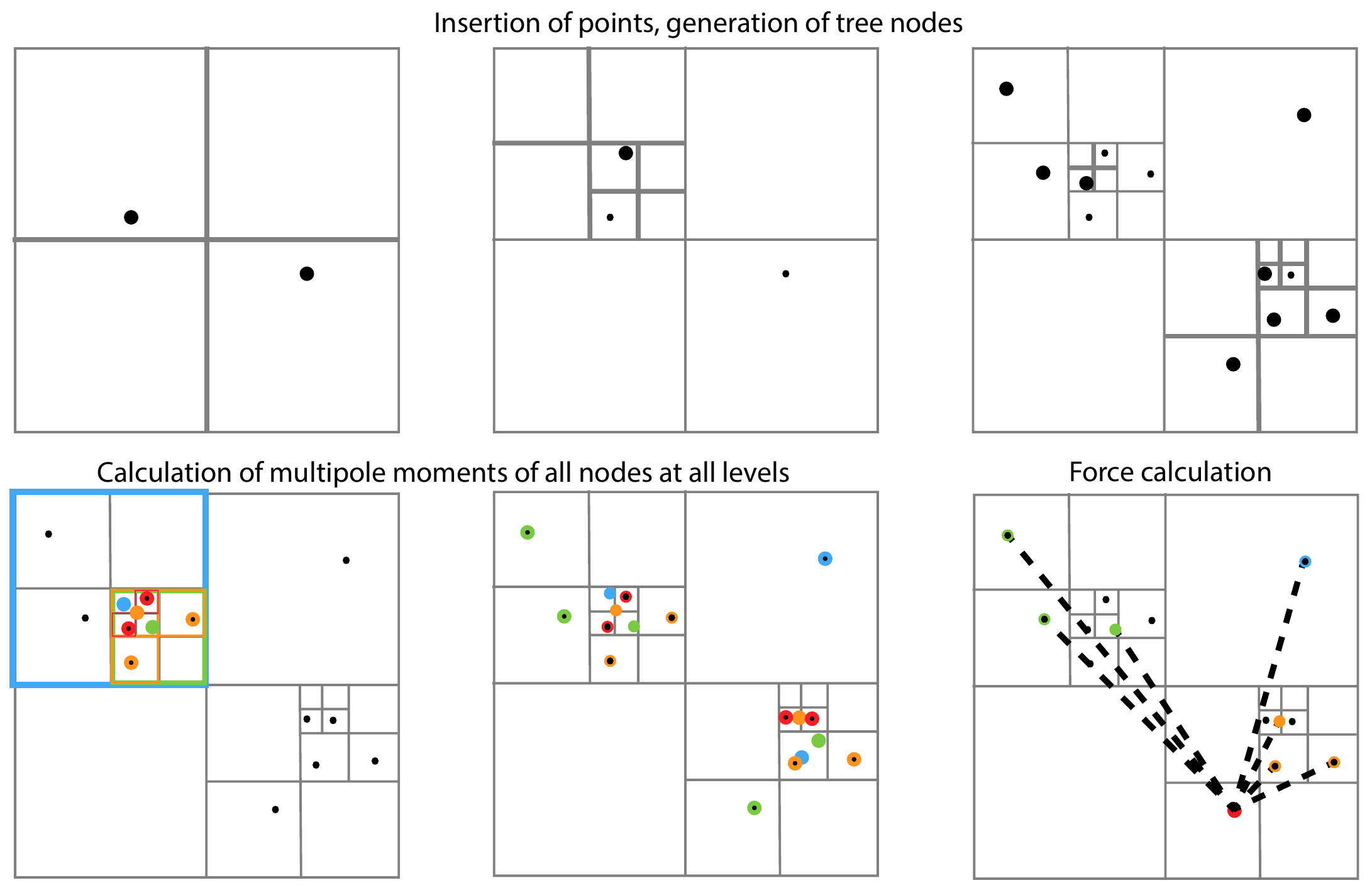}
  \caption{{Illustration of tree construction, multipole-moment calculation,
  and force evaluation of one individual particle in an oct-tree algorithm 
  (a quad-tree in the sketched 2D case). The upper panels illustrate the sequential
  insertion of simulation particles and the associated hierarchical splitting 
  of the tree volume into sub-nodes. In the lower left and central panels, 
  the calculation of multipole moments of all nodes is illustrated, with 
  different colors denoting different depths of the tree hierarchy. Finally, the 
  lower right panel illustrates the force evaluation between a single particle 
  and the appropriate nodes. Reducing the number of these evaluations 
  with respect to an evaluation with all simulation particles is the goal of 
  this algorithm.}}
  \label{fig:oct-tree}
\end{figure}

The Barnes-Hut \citep{Barnes+1986} oct-tree algorithm starts with an
all-enclosing root node, which is then split up recursively into eight
sub-nodes of equal volume, provided there are still particles
contained in these volumes ({see Figure~\ref{fig:oct-tree}}). 
The version implemented in \textsc{Arepo}
is similar to the one presented in \citet{Springel2005}; in particular,
only monopole moments\footnote{Dipole moments are implicitly included
  as well because the center of mass of the nodes is used as the expansion
  center.} are calculated for each node, and the parallelization
strategy for walking the tree is similar, too. We note, however, that
the tree structure in \textsc{Arepo} is newly constructed for each
local time step instead of using dynamic tree updates as in
\citet{Springel2005}. This introduces additional computational cost,
but avoids subtle correlations of the force accuracy with the
time stepping hierarchy, and reduces force errors in the centers of
halos in case the tree has not been reconstructed for a long time.

The tree construction starts with an enclosing cubical root note in
which the particles are inserted sequentially one by one. Each time a
particle ends up in an already occupied leaf node, i.e.~a node with a
particle in it and without further sub-nodes, the corresponding
sub-nodes are created and the particles are distributed in the sub-nodes
recursively until all leaf nodes contain at most one particle. If
there are two particles with (almost) identical positions, this can
lead to a very deep hierarchy. To avoid this, the ordinary hierarchy
is limited to a maximum level (by default 30), at which point one of
the two particles is assigned to a randomly selected neighboring
sub-node.

For each tree node, multipole moments are computed recursively and
then used to approximate the gravitational forces.  To this end, the
tree is walked for each particle, starting at the root node. If the
multipole expansion of a visited node is considered sufficiently
accurate, it is evaluated and added to the accumulated
force. Otherwise, the node is opened and the daughter nodes are
considered in turn.  Following \citet{Springel2005}, a relative
opening criterion is normally used where a node is opened if
\begin{align}
  \frac{G m}{d^2} \frac{L_\text{node}^2}{d^2} > \alpha \left| \vec{a}\right|.
\end{align}
Here $m$ is the mass of the node, $d$ is the distance of
the particle to the node's center of mass, and
$ \left| \vec{a}\right|$ is the absolute value of the acceleration,
estimated from the previous time step, {and $\alpha$ is the code parameter 
to control the opening criterion.} 
This limits the error from every individual particle-node interaction to 
a small fraction of the total
force, so that a roughly constant relative force error can be
expected. In addition, a node is always opened if a particle lies
within a box of size 1.2 times the node side length $L_\text{node}$,
centered on the geometric center of the node. This guards against the
possibility of unusually large force errors if a particle lies close
to or within a node, but happens to be still relatively far away from
the center of mass of the material in the box \citep{Salmon+1994}.

A classic  geometrical criterion
\begin{align}
  L_\text{node} > d \, \vartheta_\text{opening}
\end{align}
may also be used instead of the relative criterion, where the opening
angle $\vartheta_\text{opening}$ is a parameter effectively
controlling the resulting force accuracy.  However, this criterion
tends to be more costly at equal force accuracy than the relative
criterion, making the latter the preferred choice. For the very first
force calculation, the geometrical criterion is always used because no
estimate of the absolute size of the force is available yet. If the
relative criterion is in use, the force is then recomputed with the
relative criterion.

Normally, the oct-tree algorithm takes into account the gravitational
interactions of all particles in the simulation volume using a
Newtonian force law, implying nonperiodic boundary conditions. If
one, however, wants to compute the forces assuming periodic boundary
conditions, the contributions of (infinitely many) mirrored images need
to be summed up as well. To achieve the latter, the Ewald summation
technique for cosmological simulations presented in
\citet{Hernquist+1991} is used.

For isolated self-gravitating systems at rest with respect to the
simulation coordinate system, the opening of nodes will be similar
between different time steps. This means that the force errors made due
to the approximation of the force via the tree are correlated at
different times. Because all force errors in the tree algorithm do not
necessarily add up to zero, this can lead to a net force on the
system, and thus a nonconservation of momentum of the entire system
that builds up with time. 

To reduce this effect, \textsc{Arepo} randomizes the placement of the
tree domain center for each tree construction for simulations with
nonperiodic boundary conditions for gravity, which de-correlates the
force errors in time and greatly improves the global momentum
conservation. For simulations with periodic boundary conditions,
\textsc{Arepo} instead shifts the whole box by a new random vector in
every domain decomposition. This shift is done for all of the coordinate
variables used in the code and is transparent to input and output.

\subsection{Gravitational softening}

When sampling continuous phase-space distribution functions with
discrete particles, it is important to realize that the direct
interaction between close particles would introduce artificial
two-body interactions that violate the premise of collisionless
dynamics.  To avoid this, a gravitational force softening is
introduced in $N$-body simulations of collisionless systems, which
replaces the Newtonian force below a softening scale $\epsilon$ with a
reduced force that smoothly declines to zero for vanishing
distance. \textsc{Arepo} employs the same softening law as the \textsc{Gadget}
code, where the potential of a point mass at zero lag is
$-G m/ \epsilon$. The force becomes fully Newtonian at $2.8\,\epsilon$,
but strong reductions of the force due to softening only occur for
$d \sim \epsilon$ and smaller distances. The force softening de facto
also limits the maximum acceleration one particle can cause,
simplifying the orbit integration, and it can protect against the
formation of bound particle pairs. A detailed analysis of the optimum
softening value in the context of cosmic structure formation is
presented in \citet{Power+2003}.

\textsc{Arepo} allows collisionless particles to have different
softening lengths, either assigned through a concept of `particle
types', where the softening of each type can be set separately, or
through properties of individual particles, such as the particle mass.
For gas cells, the gravitational softening length is always chosen
in a variable way,
\begin{align}
  \epsilon_\text{cell} = f_h \left(\frac{3 \, V}{4 \, \pi} \right)^{1/3},
\end{align}
where $V$ is the volume of the Voronoi cell and $f_h$ is an input
parameter that controls the size of the softening in relation to the
cell size.

Interactions between particles with different softening lengths are
symmetrized by adopting the larger of the two. If a node contains
particles with different softening lengths, and the target particle's
softening is smaller than the node's maximum softening length, and the
distance to the node's center of mass is smaller than the node's
maximum softening, a node is always opened, because otherwise the
multipole expansion may effectively account for some interactions with
the wrong symmetrized softening. Otherwise, the node may be used if
permitted by the opening criteria, using a multipole expansion of the
softened interaction potential where appropriate.

\subsection{Particle-mesh algorithm}

Alternatively to calculating individual particle interactions, it is
also possible to bin the mass distribution on a Cartesian, regular
grid via cloud-in-cell (CIC) assignment, solve for the gravitational
potential on this grid, finite difference it to get the force, and
then interpolate the force field to the particle positions. A
conceptionally simple way of doing this in the periodic case is to use a
discrete Fourier transformation to convert the Poisson Equation
(\ref{eq:Poisson}) to
\begin{align}
 - \vec{k}^2 \Phi_k = 4 \pi G \rho_k.
\end{align}
Hence a simple division of the Fourier-transformed density field
$\rho_k$ with the square of the wavevector $\vec{k}$ is sufficient to
obtain the Fourier-transformed gravitational potential. Using an
inverse Fourier transform then yields the potential field, which
subsequently can be differenced and interpolated to the particle
position. In practice, we apply an additional deconvolution operator
to correct for the smoothing effects of the CIC assignment and the
tri-linear interpolation \citep{Hockney1981}.

\textsc{Arepo} allows for the use of a mesh covering the entire
computational domain to calculate large-scale forces
(\texttt{PMGRID}\footnote{Here and in the following, we point out 
compile-time options of the code that are controlling features discussed 
in the text, for easier reference.} option). This is commonly applied for
cosmological volume simulations, where this approach also conveniently 
yields periodic boundary conditions. However, the code can also compute 
mesh-based forces for nonperiodic boundaries. In this case, zero-padding 
and a different Green's function in Fourier space is used.  For zoom
simulations, it is additionally possible to place a second
(nonperiodic) mesh onto the usually small high-resolution region in order to
extend the dynamic range covered with this technique
(\texttt{PLACEHIGHRESREGION}). For these types of simulations, an
alternative communication algorithm for binning the density field and
interpolating the forces (\texttt{PM\_ZOOM\_OPTIMIZED}) is also
available, allowing reasonable workload balance in the PM
calculations even for highly inhomogeneous particle distributions.
{The particle-mesh algorithm makes use of the 
one-dimensional fast Fourier transformation of the FFTW library.
The use of external libraries is discussed in more detail in
Section~\ref{sec:usage_of_libraries}. }

\subsection{Tree-particle-mesh approach}

Both the tree and the particle-mesh methods have advantages and
disadvantages. The particle-mesh algorithm is conceptionally simple,
fast, and comparably easy to scale to a large number of massage passing 
interface (MPI) tasks.
However, it is severely limited by being bound to a uniform Cartesian
mesh, limiting the dynamic range of the calculation
significantly. This poses a severe obstacle for cosmological
simulations of galaxy formation. On the other hand, the particle-mesh
algorithm yields periodic boundary conditions in a computationally
highly efficient and accurate way, which is perfectly suited for
cosmological simulations. 

The tree algorithm in its standard form assumes nonperiodic boundary
conditions, and extending this to periodic boundaries is
computationally costly, especially for the small density perturbations
found on an otherwise homogenous background at high redshift.  On the
other hand, it can naturally and efficiently handle large dynamic
ranges in spatial scales, allowing very high spatial force
resolutions. In addition, it can be easily combined with local and
adaptive time stepping.  One disadvantage however is that scaling to a
large number of MPI tasks is more difficult, especially if the
particle distribution is highly clustered.

To benefit from the advantages of both methods, \textsc{Arepo} is able
to split forces into short-range and long-range components and compute
the former with its tree algorithm and the latter using the particle-mesh 
approach.  The resulting TreePM algorithm \citep{Bagla2002} is
implemented in a conceptually similar way as in \textsc{Gadget}
\citep{Springel2005}.  The force is split into a long-range and a
short-range contribution, where the short-range force of  a point mass
is obtained from the Newtonian force by multiplying 
by
\begin{align}
 f_l = 1 - \text{erfc}\left(\frac{r}{2 r_s}\right) - \frac{r}{\sqrt{\pi} r_s} \exp \left( \frac{r^2}{4 \, r_s^2} \right).
\end{align}
In Fourier space, the complementary long-range force is obtained
from the long-range potential 
\begin{align}
  \Phi_\text{k, long} = \Phi_\text{k} \exp\left( - \vec{k}^2 r_s^2 \right)
\end{align}
where the split scale $r_s$ is specified in terms of a factor  $a_s$,
giving this length in units of the PM mesh cell size,~i.e.,
\begin{align}
  r_s &= a_s \frac{L_\text{box}}{N_\text{pm}}.
\end{align}
When walking the tree, nodes or particles farther away than a cutoff
radius $a_\text{cut} \, r_s$ are simply ignored and need not be
evaluated, because $f_l$ has dropped to a negligible value there and
the corresponding force is provided by the PM algorithm. The ability
to discard the mass distribution except for a local neighborhood then
accelerates the tree force calculation significantly, making TreePM
ultimately a fast, accurate, and still very flexible force calculation
approach.

For cosmological simulations, the code allows us to have two layers of
particle-mesh calculations, as described above, the first covering the
complete box, the second only the high-resolution region. The tree
walk is then able to use a shorter cutoff length in the high-resolution 
region than in the more coarsely sampled boundary region of
low resolution. The standard parallel fast Fourier transform (FFT) package used by
\textsc{Arepo} supports parallel Fourier transforms with a slab-based
decomposition. In this case, the mesh size directly limits the number
of MPI ranks that can be efficiently used, which can become a
restriction for very large simulations. To avoid this limitation,
\textsc{Arepo} offers alternatively a column-based FFT 
(\texttt{FFT\_COLUMN\_BASED}). The corresponding algorithm requires 
more transpose operations and is therefore more costly for small 
transforms, but its better scalability ultimately allows larger maximum 
transform sizes. In Table~\ref{TabGravParams} we summarize the 
names of code parameters affecting the gravity calculation, and give their 
mapping to the symbols used in the above equations.

\begin{table}
\caption{Code parameters for gravitational force calculation.  \label{TabGravParams}} 
\centering
{
\begin{tabular}{|l|c|c|}
  \hline
  Description & Symbol & Code Parameter \\
  \hline \hline
  Tree opening angle & $\vartheta_\text{opening}$ & \texttt{ErrTolTheta} \\
  \hline
  Acceleration opening criterion & $\alpha$ & \texttt{ErrTolForceAcc} \\
  \hline
  Gas softening length relative to cell radius & $f_h$ & \texttt{GasSoftFactor}\\
  \hline
  Force split scale & $a_s$ & \texttt{ASMTH} \\
  \hline
  Short-range force cutoff & $a_\text{cut}$  & \texttt{RCUT} \\
  \hline
\end{tabular}
}
\end{table}

\section{Discretization of MHD}
\label{sec:hydro}

To solve the equations of (magneto)hydrodynamics, \textsc{Arepo} uses a
second-order accurate finite-volume discretization. To this end,
volume-averaged primitive variables $\rho$, $\vec{v}$ and $\vec{B}$
are stored as properties of the cell at its center. Gradients are
estimated with the corresponding values for neighboring cells
\citep{Pakmor+2016}, allowing for a piecewise-linear reconstruction of the
solution. Using the gradients, the primitive variables are
extrapolated to all mesh interfaces, for which fluxes are calculated by
solving a Riemann problem locally at each interface. The flux
calculation can be done either with an exact, iterative Riemann solver (default)
or the approximate HLLC (\texttt{RIEMANN\_HLLC}) solver in the case of pure
hydrodynamics and an HLLD (\texttt{RIEMANN\_HLLD}) solver in the case of 
MHD \citep[see ][for details]{Toro1997}.

\subsection{Gradient estimate}

As shown in \citet{Pakmor+2016}, the gradient estimate of the
hydrodynamic quantities originally presented in \citet{Springel2010},
based on a finite difference formula for Voronoi cells, can become
inaccurate for highly distorted cells. The present version of
\textsc{Arepo} therefore employs an improved, least-square
gradient estimate  \citep{Pakmor+2016}. The basic idea is that the
gradient $\nabla W_i$ of a primitive variable $W_i$ is determined such
that the linearly 
extrapolated values 
from cell $i$ to the positions of the neighboring cells $j$, 
\begin{align}
  \tilde{W}_j = W_i + \vec{d}_{ij}\, \nabla W_i ,
\end{align}
agree with the actual values $W_j$ found there as well as possible.  Here
$ \vec{d}_{ij}$ is the position vector of cell $j$ relative to cell
$i$. Because there are multiple neighbors, exact equality can
generally not be achieved; rather, we determine a gradient estimate by
minimizing the residuals of the over-determined set of equations in a
weighted least-square sense. Specifically, we minimize
\begin{align}
  S_\text{tot} = \sum\limits_j g_j \left( W_j - W_i - \vec{d}_{ij}\,\nabla W_i \right)^2.
\end{align}
The adopted weights are given by
$g_j = A_{ij} / \left|\vec{d}_{ij}\right|$, where $A_{ij}$ is the area
of the interface between $i$ and $j$.

\subsection{Divergence constraint}

In their analytical form, the equations of MHD
automatically conserve the $\nabla \cdot \vec{B} = 0$ constraint
provided that it is fulfilled initially. This property, however, is
generally not true for discretizations of these equations \citep[see
however][]{Evans+1988}. In this version of \textsc{Arepo}, we adopt
the divergence-cleaning method introduced by \citet{Powell+1999}, as
implemented into the code by \citet{Pakmor+2013}. This method advects
numerically induced divergences away, and has the advantage of using
cell-centered magnetic fields. Thus, the momentum, energy, and induction
equations (\ref{eq:euler:momentum} - \ref{eq:euler:induction}) become
\begin{align}
  \frac{\partial \rho \vec{w} }{\partial t} &+ \nabla \cdot \left( \rho \vec{u} \vec{u}^{T} + p_\text{tot} - \frac{\vec{B} \vec{B}^{T}}{a} \right) = -\frac{1}{a} \left( \nabla \cdot \vec{B} \right) \vec{B} .\\
  \frac{d\mathcal{E} }{dt} &+ a \nabla \cdot \left[ \vec{u} \left( E + p_\text{tot} \right) - \frac{1}{a} \vec{B} \left( \vec{u} \cdot \vec{B} \right) \right] = \frac{\dot{a}}{2} \vec{B}^2 - \left( \nabla \cdot \vec{B} \right) \left(\vec{u} \cdot\vec{B} \right),\\
  \frac{\partial \vec{B}}{\partial t} &+ \frac{1}{a} \nabla \cdot \left( \vec{B} \vec{u}^{T} - \vec{u} \vec{B}^{T} \right) = -\frac{1}{a} \left( \nabla \cdot \vec{B} \right) \vec{u}.
\end{align}
The divergence of the magnetic field in a cell $i$ is calculated as
\begin{align}
  \nabla \cdot \vec{B}_{i} = V_i^{-1} \sum\limits_\text{faces} \vec{B}_\text{face} \cdot \hat{\vec{n}} \, A_\text{face},
\end{align}
with $\hat{\vec{n}}$ being the normal vector to the cell,
$\vec{B}_\text{face}$ is the value of the magnetic field on the
interface, and $V_i$ is the volume of the cell. The advantage of this
method is that it is very flexible in terms of mesh geometry, and the
possibility to discretize it on local time steps makes it computationally
inexpensive {and thus suitable for use in cosmological 
simulations of galaxy formation. But the 8-wave formulation has known 
shortcomings, e.g. in reproducing jumps in an inclined 2D MHD shocktube test 
\citep{Toth2000, Mignone+2010}, and \textsc{Arepo} does not reach as low
numerical reconnection rates as state-of-the-art Cartesian grid MHD codes 
(see magnetic current-sheet example, section~\ref{sec:current_sheet_2d}).} 

\subsection{Gravitational interactions of the fluid}

To compute the gravitational forces of the fluid, the gas cells are
included in the gravitational force calculation as described above.
The code uses cell-centered softened gravitational forces with  softening
lengths tied to the cell radii, such that the gravitational field
created by the gas distribution is a good approximation 
of a space-filling continuous density distribution.

The gravitational forces are coupled as source terms to the MHD
equations, in an operator split approach. This means the source terms
are applied for half a time step at the beginning of a step, then the
hyperbolic set of fluid equations are evolved for one time step in
their conservative form, followed by another gravity half-step. In the
case of a self-gravitating fluid, this approach does not manifestly
conserve total energy, which is difficult albeit not impossible to
achieve for mesh codes \citep{Jiang2013}. However, errors in total
energy quickly decrease for better resolution, and the source term
approach can avoid unphysical drifts in thermal energy in poorly
resolved flows for conservative treatments \citep[see also the
discussion in Section~5 of][]{Springel2010}; hence, we prefer it in
practice. For simplicity, we refrain from including correction factors
in the forces due to changes in the gravitational softening
lengths. They were still included in \citet{Springel2010} but are
typically quite small in practical applications. And as they can act
as an additional source of noise, it is unclear whether they are
ultimately beneficial for the final accuracy.

\section{Computational mesh}
\label{sec:mesh}

\begin{figure}
\centering
  \includegraphics[width=1.0\textwidth]{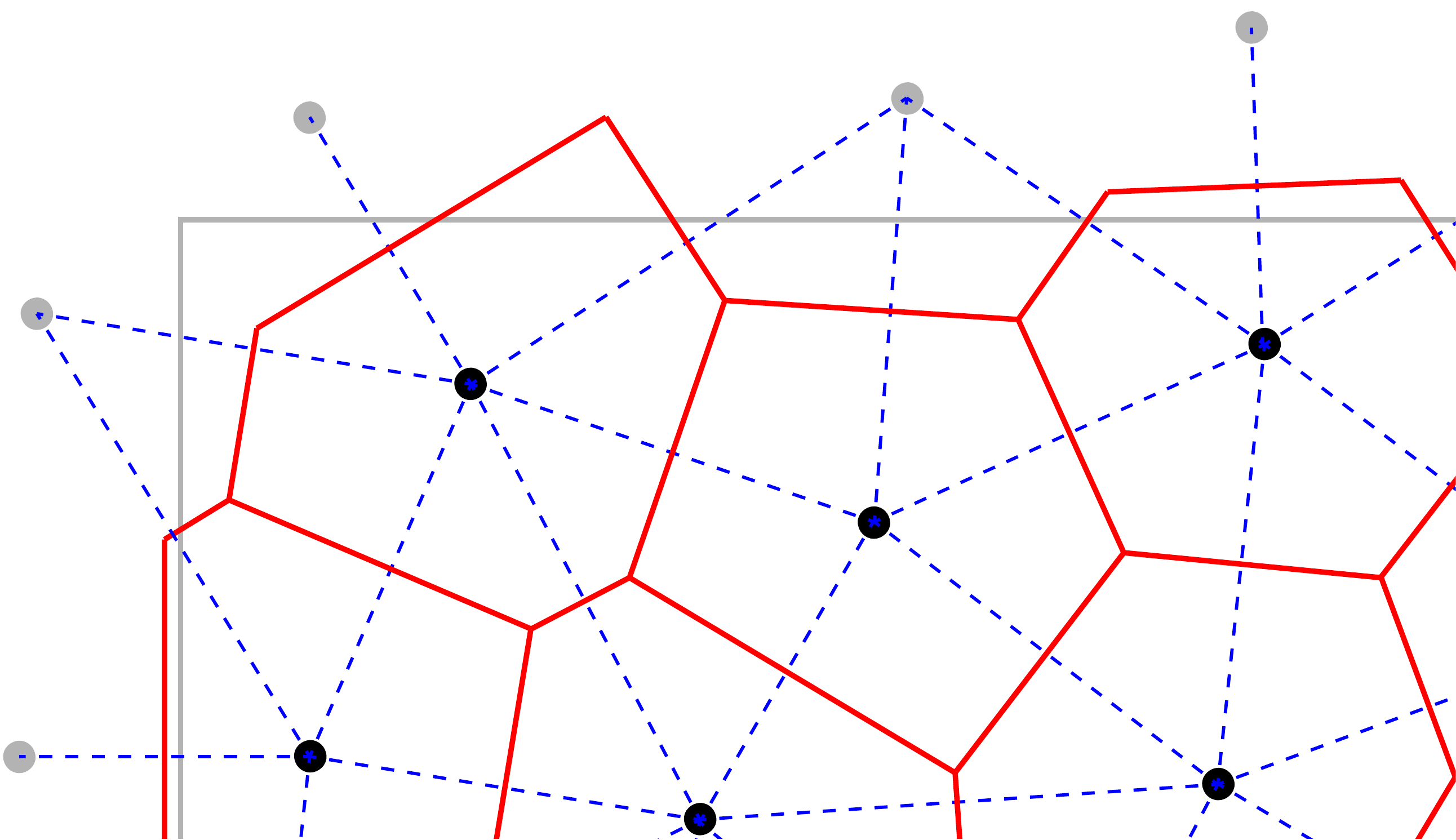}
  \caption{{Schematics of the Voronoi mesh in a single domain
  and its connections to neighboring domains. The domain boundary is 
  denoted in gray, the Delaunay connections in  and the Voronoi mesh
  in red. The mesh-generating points of the local domain are shown 
  in black, and the ones of different domains, imported as ghost cells, are 
  drawn as gray dots. Which domain a given cell is associated with
  is determined by the location of the mesh-generating point, not by
  the cell boundary. Unlike in a fixed grid-geometry, ghost points from 
  other domains (or boundaries) need to be imported to determine the 
  local mesh.}}
  \label{fig:mesh}
\end{figure}

A characteristic feature of the \textsc{Arepo} code is its
discretization on a fully adaptive, dynamic Voronoi mesh. The starting
point for creating this Voronoi mesh are mesh-generating points, each
one representing the position of a Voronoi cell. Each Voronoi cell is
defined as the region in space that is closer to a given
mesh-generating point than to any other mesh-generating point. This
implies that mesh interfaces are normal to the connection of two
neighboring cells, and lie at equal distances to the mesh-generating
points on either side.

The ensemble of connections to neighboring cells (and only to them)
forms a so-called Delaunay triangulation, consisting of tetrahedra in
three dimensions. The circumsphere of each tetrahedron does not
contain any other mesh-generating point, and it is this mathematical
property that makes the Delaunay tessellation special and unique among
all other possible triangulations of space.  The Delaunay tessellation
is computed iteratively in \textsc{Arepo}: we start with four points
(in three dimensions) that span a domain-enclosing tetrahedron.  Now,
we successively insert the mesh-generating points into the Delaunay
mesh, adding connections from the point to the corners of the
tetrahedron the point falls in. This splits the tetrahedron into four
smaller tetrahedra. However, each of the resulting tetrahedra is not
necessarily a Delaunay tetrahedron, i.e., it may violate the empty
circum-circle property.  Delaunayhood is, in this case, restored by
iteratively applying edge-flipping to affected tetrahedra
\citep[see][for a detailed discussion of all possible
cases]{Springel2010} before the next point is inserted. Once the
Delaunay triangulation is completed, all mesh-generating points are
part of the Delaunay mesh, and the Voronoi mesh is trivially obtained
as its topological dual, requiring only simple geometrical
calculations such as the areas of the interfaces. 

In simulations with distributed-memory parallelization, the above
algorithm is first used to compute local meshes for each domain, which
are then complemented at the borders by inserting further mesh-generating 
points imported as `ghost points' from neighboring
domains. In this way, Voronoi cells that overlap with domain
boundaries are still constructed with their correct shape.
{An illustration of the mesh in a single domain, and its connections
to points from neighboring domains, is shown in Figure~\ref{fig:mesh}.
How these domains are determined is discussed in 
section~\ref{sec:loadbalancing}.}

\subsection{Mesh movement}

Each mesh-generating point normally moves with the bulk velocity of
the fluid in its cell, also taking into account accelerations due to
local pressure gradients, the Lorenz force, and gravity, i.e., de facto
adopting the expected velocity at the next half time step. This
naturally leads to an approximate equal-mass discretization and to
quasi-Lagrangian behavior. This, however, does not ensure by itself
that the mesh remains reasonably regular at all times: due to local
fluid motions, high aspect ratios of mesh cells can occur, which
increase discretization errors and mesh noise, and may thus negatively
impact the accuracy of the solution \citep[as discussed
in][]{Springel2010, Duffell+2011, Vogelsberger+2012}.

To avoid this, velocity corrections can be applied for highly
distorted cells to steer the mesh motion \allowbreak
(\texttt{REGULARIZE\_MESH\_CM\_DRIFT}). 
We identify highly distorted
cells by the maximum angle under which any of its faces is seen from 
the mesh-generating point
(\texttt{REGULARIZE\_MESH\_FACE\_ANGLE}), i.e.,
\begin{align}
  \alpha_\text{face} = (A_\text{face} / \pi)^{1/2} / h_\text{face}, \\
  \alpha_\text{max} = \max\left( \alpha_\text{face} \right),
\end{align}
where $A_\text{face}$ is the area of a given interface and
$h_\text{face}$ is the distance from the mesh-generating point to the
interface. If $\alpha_\text{max}$ exceeds a predefined threshold
$0.75\,\beta$, with $\beta$ being a free parameter, the
mesh-generating point is moved with a fraction $f_\text{shaping}$ of
the characteristic speed in the cell toward its center of mass.
The corrective velocity is parameterized as:
\begin{align}
 \vec{v}_\text{corr} =
 \begin{cases}
 0 \qquad & \text{ for } \alpha_\text{max} \leq 0.75\,\beta \\
 f_\text{shaping} \, \frac{\alpha_\text{max} - 0.75\, \beta}{0.25\, \beta} v_\text{char} \, \hat{\vec{n}} \qquad & \text{ for } 0.75\,\beta < \alpha_\text{max} \leq \beta \\
 f_\text{shaping} \, v_\text{char} \, \hat{\vec{n}} \qquad & \text{ for } \alpha_\text{max} >  \beta \\
 \end{cases}
\end{align}
\citet[][section 2.2.2 (i)]{Vogelsberger+2012} discusses the use of
this mesh regularization approach in the context of cosmological
simulations. It has become the default for these kinds of simulations
since.  However, the modified Lloyd scheme presented in
\citet{Springel2010} can also be used as an alternative
(\texttt{REGULARIZE\_MESH\_FACE\_ANGLE} not set), for which the
correction velocity is given by
\begin{align}
 \vec{v}_\text{corr} =
 \begin{cases}
 0 \qquad & \text{ for } d \leq 0.75 \, \eta \, r_\text{cell} \\
 f_\text{shaping} \, \frac{d - 0.75 \, \eta \,r_\text{cell}}{0.25 \, \eta \, r_\text{cell}}\, v_\text{char} \, \hat{\vec{n}} & \text{ for } 0.75 \, \eta \, r_\text{cell} < d \leq \eta \, r_\text{cell} \\
 f_\text{shaping} \, v_\text{char} \, \hat{\vec{n}} \qquad & \text{ for } d > \eta \, r_\text{cell} \\
 \end{cases}
\end{align}
with $d$ being the distance between the mesh-generating point and center
of mass of a cell. For the characteristic speed $v_\text{char}$ of a
cell, the sound speed can be used
(\texttt{REGULARIZE\_MESH\_CM\_DRIFT\_USE\_SOUNDSPEED}), or
$d/\Delta t$ where $\Delta t$ is the time step. In the latter case,
$v_\text{char}$ is at most equal to the maximum hydrodynamic velocity
in the system.  Table~\ref{TabMeshParams} summarizes the most
important run-time parameter of the code affecting the hydrodynamics
and the mesh motion.

\subsection{Refinement and de-refinement}
 
Even when using quasi-Lagrangian mesh motion, over the course of many
simulation time steps, cells can diverge from their desired mass
content or intended size. For this reason, \textsc{Arepo} offers the
possibility to refine and de-refine the mesh locally, thus offering
full spatial adaptivity\footnote{{Note that the refinement/de-refinement operations are local
to a given cell and do not trigger any global load-balancing operations.}}.

To refine a cell, the mesh-generating point is split into a pair of
two very close points,
offsetting 
their location in a random direction by $0.025\, r_\text{cell}$, where
\begin{align}
  r_\text{cell} = \left( \frac{3 V}{4 \pi} \right)^{1/3}.
\end{align}
This effectively splits the original Voronoi cell into two cells,
without affecting the geometry of the neighboring cells.  The
conserved quantities of the split cell are then subdivided
conservatively among the two new cells according to their volume
ratio.

For de-refinement, a mesh-generating point is taken out, thereby removing
the corresponding Voronoi cell from the mesh, with the neighboring
cells claiming its volume. The conserved quantities associated with
the eliminated cell are conservatively distributed among these neighbors,
in proportion to the volume overlap they realize with the removed
cell.

\begin{table}
\caption{Code parameters for hydrodynamics and mesh movement. \label{TabMeshParams}}
\centering
{
\begin{tabular}{|l|c|c|}
  \hline
  Description & Symbol & Code Parameter \\
  \hline \hline
   Adiabatic index & $\gamma$ & \texttt{GAMMA} \\
  \hline
  Cell roundness criterion & $\beta$  & \texttt{CellMaxAngleFactor} \\
  \hline
  Alternative cell roundness criterion & $\eta$ & \texttt{CellShapingFactor} \\
  \hline
  Cell deformability parameter & $f_\text{shaping}$ & \texttt{CellShapingSpeed} \\
  \hline
\end{tabular}
}
\end{table}

In principle, nearly arbitrary refinement criteria are possible, and
very different ones have been employed in certain simulations
\citep[e.g.][]{vandeVoort+2019}. In the public version, we include an
often used mass criterion, which triggers de-refinement if a cell has
less than half of a desired target mass resolution, and refinement if
it has more than twice this target mass. This keeps the mass
resolution in a narrow corridor around the target mass resolution,
thus realizing a Lagrangian behavior akin to smoothed particle 
hydrodynamics (SPH) codes.  Additionally,
a refinement criterion based on resolving the Jeans length by a given
number of cells can be employed. This is usually used to ensure that
gravitational collapse and fragmentation are resolved sufficiently
well, and not significantly impacted by resolution issues.

We note that normally highly distorted cells are excluded from
refinement, in order to avoid that refinements triggered in quick succession in
subsequent time steps produce locally a very irregular mesh. After a
refinement event, the mesh steering motions take a couple of time steps
to reestablish a locally regular mesh geometry, at which point the next
refinement may then proceed.

\subsection{Boundary conditions}
For gravity, \textsc{Arepo} supports either periodic or nonperiodic
(vacuum) boundary conditions. For hydrodynamics, periodic boundary
conditions are one of the primary possibilities and represent the
simplest choice, but the code also supports reflective and in-/outflow
boundary conditions at the box borders. Differently from SPH and pure
particle-based gravity, the confines of the simulated volume need to
be unambiguously specified in \textsc{Arepo}. At least for
hydrodynamics, a box size must therefore always be specified.

Reflective boundaries are effectively realized by mirroring the
mesh-generating point set at the box border, thereby ensuring that
Voronoi faces align with the box boundary. The fluid state is then
also mirrored, so that the Riemann solvers return zero mass flux at
the corresponding box border. Inflow or outflow boundaries are
realized similarly, except that the fluid state is not simply mirrored
at the interface but replaced, for example, with a predefined state
describing the inflow conditions into the simulated volume. For
hydrodynamics, periodic, reflective, or inflow/outflow boundaries can
be independently selected for each spatial dimension.

\section{Additional physics}
\label{sec:physics}

For a number of astrophysical systems, modeling just gravity and MHD 
already provides an interesting approximation that warrants detailed
study. However, there are a number of problems where this alone is
only a starting point and further important physics needs to be added.
This can, for example, be introduced through source or sink terms of
thermal energy due to interaction with radiation fields or local
energy production in nuclear reactions. As the precise nature, and
thus the implementation, of such terms depends highly on the system
under study, and the verification of the corresponding calculations is
far less clear than in the gravity+MHD case, we include only two very
simple modules for such extra physics here to provide an illustrative
guideline for how more sophisticated modules can be implemented.

The first example is a sink term due to radiative cooling, and the
second is a simple sub-grid model for the interstellar medium and its
embedded star formation. Both of these models have been extensively
applied in the literature to simulations of galaxy formation, but
recent work has largely moved on to more sophisticated treatments.
All equations in the following section are in proper, not co-moving
coordinates, and are also implemented in this way in the code.

\subsection{Radiative cooling}

One of the key additional physics ingredients to an optically thin
astrophysical plasma is radiative cooling. In \textsc{Arepo}, this is
implemented as a sink term in the energy equation as described in
equation (\ref{eq:euler:energy}). The main challenge with radiative
cooling is that this loss term is a strong function of metallicity and
density, making the cooling rate a stiff equation that can be
difficult to integrate numerically. In order to avoid tight time step
constraints that an explicit integration scheme for radiative cooling
would impose, \textsc{Arepo} uses a {first-order} implicit integration of
the cooling term, based on a root-finding algorithm of the equation
\begin{align}
  u^\prime - u_n -\frac{\Lambda(u^\prime)}{\rho} \Delta t = 0.
 \end{align}
 The solution for $u^\prime$ at the end of a step is found iteratively, evaluating the
 cooling function $\Lambda$ at the respective specific internal energy
 and ionization state for the next iteration step. In the public
 version of the code, only a primordial cooling network for H and He
 \citep{Cen1992,Katz+1996} under the assumption of collisionless
 ionization equilibrium is included; however, extensions to this model are
 straightforwardly possible.

\subsection{Modeling star formation}

Upon including gas cooling in hydrodynamical simulations of galaxy
formation, one quickly arrives at situations where the gas looses its
pressure support and collapses under the relentless pull of
gravity. In reality, this gravitational collapse is very complex and
occurs over roughly $10$ orders of magnitude before local fragments
give birth to individual stars. In simulations, however, it is both
impractical and currently impossible to follow the gravitational
collapse of a molecular cloud to all of the stars it forms and the onset of
nuclear reactions in them, let alone in an entire galaxy. Instead, one
needs to resort to much more simplified treatments, taking the form of
so-called sub-grid models. They stand for the general idea of
introducing an effective model for complex unresolved astrophysical
processes, which is often necessary when the physics cannot be
resolved directly, but it comes at the price of adding uncertainty and
heuristic input.

For example, one can define, as is commonly done, a density threshold
above which gas forms so-called star particles with an estimated mean
rate. The star particles represent stellar populations, and they are
collectively modeled as a population of collisionless particles that,
in particular, no longer interact with the surrounding gas via
hydrodynamical forces.  The implemented example of star formation in
the public release of \textsc{Arepo} follows the \citet{Springel+2003}
model for treating the unresolved interstellar medium. We note,
however, that the metal enrichment and wind formation models
\citep[sections 5.3 and 5.4 of][]{Springel+2003} are not part of this
implementation.

The basic idea of this model is to not attempt to resolve the
multiphase gas structure in the ISM, but instead represent its
spatially averaged behavior through a simple smooth model. The ISM is
pictured as a two-phase medium in pressure equilibrium, consisting of
cold clouds in which star formation can occur and which are embedded
in a hot, space-filling medium. For a detailed discussion of the
model, we refer to \citet{Springel+2003}, and only list the primary
implemented equations and their connection to the free parameters of
the model here. In particular, the effective specific thermal energy
of a star forming cell, i.e., a cell with density exceeding
$\rho_\text{th}$, is given by
\begin{align}
  u_\text{eEoS} = u_\text{hot} (1-x) + u_\text{c} x ,
\end{align}
with $x$ being the mass fraction of cold clouds, calculated as
\begin{align}
x &= 1 + \frac{1}{2 \, y} - \sqrt{ \frac{1}{y} + \frac{1}{4 y^2} ,} \\
y &= \frac{t_*}{t_\text{cool} } \frac{ u_\text{hot}  }{\beta u_\text{SN} - (1 - \beta) u_\text{c} }.
\end{align}
Here $\beta$ is the fraction of stars according to the initial mass
function that explode as supernovae, and $t_\text{cool}$ is the
cooling time derived from the employed cooling model.  The specific
internal energies can be associated with the temperature parameters
via
\begin{align}
  u_\text{c} = \frac{k_B}{(\gamma - 1) \,\mu_\text{neutral} \, m_p} T_\text{c}, \\
  u_\text{SN} = \frac{k_B}{(\gamma - 1) \,\mu_\text{ionized} \, m_p} T_\text{SN},
\end{align}
where a completely neutral and fully ionized gas is assumed, respectively.
The specific energy in the hot ISM phase is given by
\begin{align}
  u_\text{hot}  = \frac{u_\text{SN} }{1 + A(\rho)} + u_\text{c},
\end{align}
and the supernova evaporation parameter is given by
\begin{align}
  A(\rho) = A_0 \left( \frac{\rho}{\rho_\text{th}} \right)^{-\frac{4}{5}}.
\end{align}
Finally, the star formation timescale is
\begin{align}
  t_* (\rho) = t_0^* \left( \frac{\rho}{\rho_\text{th}}
  \right)^{-\frac{1}{2} } ,
\end{align}
where star formation is only allowed to take place if the temperature is below
$T_\text{thresh}$ and the density is above $\rho_\text{th}$ and $\rho_\text{c.o.d.}$. In this
case, the star formation rate of a cell is taken to be
\begin{align}
  \dot m_* = \frac{m_\text{cell}}{t_*(\rho) },
\end{align}
and actual star particles are spawned stochastically with this
rate. If a star formation event occurs and a cell has too little mass
left afterwards, the cell is converted to a star particle in full;
otherwise, it survives with correspondingly reduced mass, energy, and
momentum content. The free parameters of this model are summarized in 
Table~\ref{TabIsmParams}.

\begin{table}
\caption{Code parameters for interstellar medium and star formation modeling.} \label{TabIsmParams}
\centering
{
\begin{tabular}{|l|c|c|}
  \hline
  Description & Symbol & Code Parameter \\
  \hline \hline
   Gas overdensity in cosmological runs above which star formation is possible & $\rho_\text{c.o.d.}$ & \texttt{CritOverDensity} \\
  \hline
   Threshold temperature below which star formation is possible& $T_\text{thresh}$ & \texttt{TemperatureThresh} \\
  \hline
   Threshold density above which model is applied & $\rho_\text{th}$ & \texttt{CritPhysDensity} \\
  \hline
   Fraction of stars that go off as supernovae & $\beta$ & \texttt{FactorSN} \\
  \hline
   Supernova evaporation parameter & $A_0$ & \texttt{FactorEVP} \\
  \hline
   Temperature of supernova in two-phase ISM model & $T_\text{SN}$ & \texttt{TempSupernova} \\
  \hline
   Temperature of cold phase of two-phase ISM model & $T_\text{c}$ & \texttt{TempClouds} \\
  \hline
   Reference star formation time ($=$ timescale at critical density) & $t_0^*$ & \texttt{MaxSfrTimescale} \\
  \hline
\end{tabular}
}
\end{table}

\section{Time integration}
\label{sec:timeintegration}

For gravity and ideal MHD, \textsc{Arepo} uses explicit
time integration, which puts certain constraints on the size of
individual local time steps both with respect to the accuracy and
stability of the scheme.  Every dark matter particle has its
own individual local gravitational time step constraint. For each cell, there are
separate time step constraints for hydrodynamics, gravitation, and
potentially for source and sink terms. Generally, the most restrictive
constraint is applied for all of the calculations of a single cell.

\subsection{Time step constraint}

For gas cells, we define a local Courant-Friedrichs-Levy (CFL) time step criterion
\begin{align}
  \Delta t \leq C_\text{CFL} \frac{r_\text{cell}}{v_\text{signal}},
\end{align}
with the Courant factor $C_\text{CFL}$ as a free parameter and the cell radius given by
\begin{align}
  r_\text{cell} = \left(\frac{3\, V}{4\, \pi} \right)^{1/3},
\end{align}
where $V$ is the volume of the Voronoi cell. The signal speed, in the
case of a moving mesh, is calculated taking into account sound speed and Alfv\'en speed
\begin{align}
  v_\text{signal} = \left(\gamma \frac{p}{\rho} + \frac{\vec{B}^2}{\rho} \right)^{1/2}.
\end{align}
In the case of a static mesh, the gas bulk velocity is added to
the signal speed, which leads to a more restrictive time step criterion
compared to the moving mesh.
For MHD, the Powell cleaning scheme imposes an additional constraint,\footnote{
{Note that formally the Powell cleaning scheme does not require additional 
time step constraints. The stated constraint is adopted to ensure
the code functionality is unaltered from the version used in many production runs.}}
\begin{align}
  \Delta t_\text{cleaning} = \frac{B + \sqrt{0.02 u_\text{th} \rho} }{2 \left| \nabla\cdot\vec{B} \right| \left| \vec{v}\right| }.
\end{align}
If the star-formation model is active, we adopt a further constraint set by
the ratio of mass of a gas cell $m$ and its star formation rate
$\dot{m}_\text{SF}$,
\begin{align}
  \Delta t_\text{SF} = 0.1 \frac{m}{\dot{m}_\text{SF}}.
\end{align}
For gravitational accelerations, we adopt the criterion
\begin{align}
  \Delta t \geq \sqrt\frac{2 C_\text{grav} \epsilon_\text{soft}}{\left| \vec{a} \right| },
\end{align}
with $C_\text{grav}$ being a free parameter.
Additionally, a nonlocal magnetohydrodynamic time step criterion
\citep{Springel2010, Gnedin+2018} is adopted  \allowbreak (\texttt{TREE\_BASED\_TIMESTEPS})
\begin{align}
  \Delta t_i \geq \min\limits_{j\neq i} \left(\frac{r_{ij}}{c_i + c_j - \vec{v}_{ij} \cdot \vec{r}_{ij}/r_{ij} } \right),
\end{align}
which is meant to preempt the earliest arrival time of hydrodynamical
waves originating in any other cell. It is implemented in practice
using a tree structure to evaluate it efficiently, avoiding the need
to loop over all particles in the time step calculation. The parameters
for the time step constraints are listed in Table~\ref{TableTimeSteps}.

\subsection{Local time stepping}

To allow for each element to be integrated with an as large as possible
but as small as necessary time step, while still maintaining a high
degree of synchronization, \textsc{Arepo} uses the common approach of
a power-of-two time step hierarchy. The total simulation time is
subdivided into $2^N$ equal steps. The corresponding time step size 
for a time bin is thus
$\Delta t = (t_\text{end}-t_\text{start})/2^N$. Particles are then
associated with the time bin that corresponds to the largest time step
that is just smaller than the most restrictive time step constraint
from gravity and MHD. A linked list of particles
belonging to each time bin is kept to be able to access particles in a
given time bin efficiently without having to search for them with a
loop over all particles. When a computational element has finished a
time step, it may change its time bin. Transitioning to a shorter
time step is always possible, but changing to a longer time step only occurs if
the current time is synchronized with the target time bin, i.e.~it is
ensured that shorter timesteps always stay nested within longer ones
throughout the whole hierarchy.

\begin{table}
\caption{Code parameters for time step constraints.}\label{TableTimeSteps}
\centering
{
\begin{tabular}{|l|c|c|}
  \hline
  Description & Symbol & Code Parameter \\
  \hline \hline
   Gravity time step parameter & $C_\text{grav}$ & \texttt{ErrTolIntAccuracy} \\
  \hline
   Courant factor for MHD & $C_\text{CFL}$ & \texttt{CourantFac} \\
  \hline
\end{tabular}
}
\end{table}

\subsection{Time stepping gravity}

The gravitational time integration is done with a second-order
accurate leapfrog scheme, expressed through alternating `drift' (which
modify the positions) and `kick' (which modify the velocities)
operations.  For fixed time step sizes, this results in a symplectic
integration scheme. The particular implementation is  very similar
to that of the \textsc{Gadget-2} code \citep{Springel2005}.

Normally, an active particle receiving a kick interacts with the full
mass distribution, independent of its time bin. This breaks manifest
momentum conservation when local time steps are used and necessitates a
full tree construction even if only a small fraction of particles
requires force calculations, negatively impacting performance for deep
time step hierarchies. This can be addressed with an alternative
hierarchical time integration approach
(\texttt{HIERARCHICAL\_GRAVITY}), in which the gravitational
Hamiltonian is systematically split, such that shorter time bins are
evolved with their own part of the Hamiltonian only. The particular
implementation of this idea as adopted in \textsc{Arepo} is based on
\citet{Pelupessy2012}, with full details given in Springel at
al.~(2020, in preparation).

\subsection{Time stepping the finite-volume scheme}

\begin{figure}
\centering
  \includegraphics[width=1.0\textwidth]{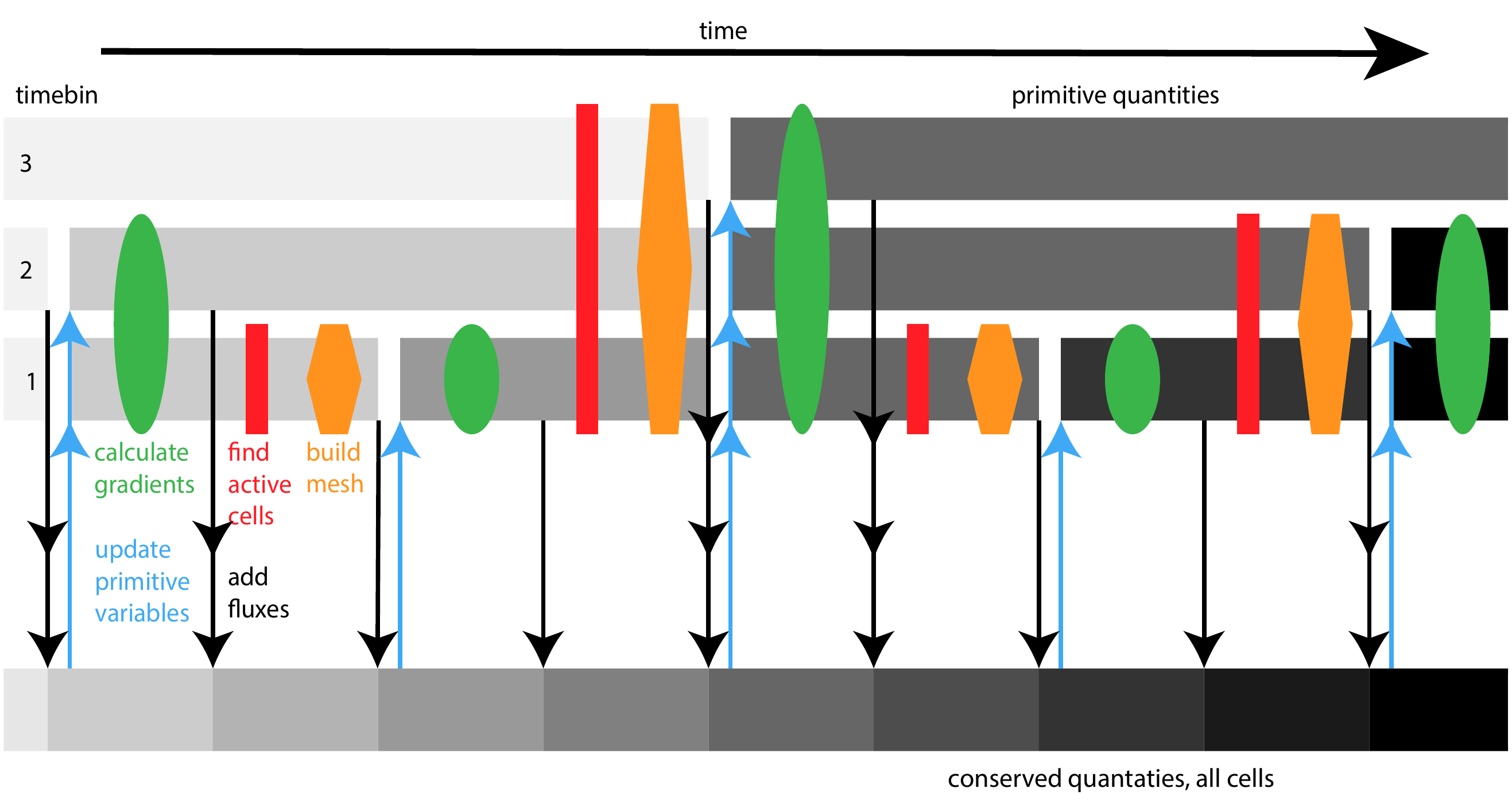}
  \caption{{Illustration of the time evolution of the Euler equations using
   local time stepping (in this case, only three time bins are sketched, for simplicity) 
   and the roles of primitive and conserved quantities. 
   For cells of each time bin, time evolution 
   follows the same steps: update of the primitive variables, calculation of 
   gradients, first flux calculation, determination of mesh at the end of the 
   time step and a second flux calculation. Fluxes through interfaces of cells
   on different time steps are added to both cells' conserved quantities during 
   the smaller time step. This leads to partially updated states in conserved quantities 
   of cells in larger time bins.}}
  \label{fig:timestepping}
\end{figure}

The time integration of the hydrodynamic quantities is described in
\citet{Pakmor+2016}, and differs slightly from the original
implementation of a MUSCL-Hancock scheme in \citet{Springel2010}. In
particular, now {a scheme more similar to} Heun's method for time 
integration is used, ensuring that only a single mesh construction is required 
for each time step while obtaining formal second-order convergence in time. 
{One key difference to static grid codes is that in a moving mesh, the facet
areas $A_{ij} = A_{ij}^n$ as well as the cell volumes $V_i = V_i^n$ are functions 
of time, and higher-order schemes need to take this time dependence 
into account.}
The update of the {volume integrated conserved variables of cell $i$, 
$Q_i = \int\limits_{\text{cell }i} U \, dV$, with
$U$ being the vector of conserved quantities, is computed as}
\begin{align}
  Q_i^{n+1} &= Q_i^n - 0.5 \Delta t \left( \sum\limits_j A^n_{ij} F(W^n_{ij}, W^n_{ji}) + A^{n+1}_{ij} F(W^\prime_{ij}, W^\prime_{ji}) \right), \quad \text{with} \\
  W^n_{ij} &= W^n_i + \vec{d}_{ij}\cdot\frac{\partial W_i}{\partial \vec{x} },\\
  W^\prime_{ij} &= W^n_i + \vec{d}_{ij}^\prime \cdot\frac{\partial W_i}{\partial \vec{x} } + \Delta t \frac{\partial W_i}{\partial t},
\end{align}
where $\vec{d}_{ij}$ is the position vector of the geometric center of the interface between cells $i$ and $j$
 relative to the center of cell $i$.  The update is thus
a combination of fluxes computed at the beginning of the current time step, and fluxes
computed on an updated mesh at the end of the time step using time-extrapolated quantities indicated by superscript $^\prime$. 
The flux calculation itself and the associated
Riemann problems are hence evaluated twice in each time step. The time
derivatives are estimated via the spatial gradients using the
continuity equation \citep[][their equations (7)-(9)]{Pakmor+2016}. 
Local time stepping is realized in a conservative
manner by adding fluxes to $Q$ in a pairwise
manner to the two cells on either side of an active interfaces,
i.e., whenever at least one of the involved cells is active
\citep[][section 7]{Springel2010} . {The time integration of 
the hydrodynamics is illustrated in Figure~\ref{fig:timestepping}.}

\section{Domain decomposition and load balancing}
\label{sec:loadbalancing}

A key feature of modern parallel simulation codes is their ability to
distribute computations over a large number of distributed-memory
compute nodes, thereby allowing very large computations on modern
supercomputers.  A prerequisite to make this possible is however that
both the computational as well as the memory load must be balanced well
across the individual nodes and processes.  To achieve this,
\textsc{Arepo} uses the MPI standard for distributed-memory
parallelization, and executes a domain decomposition routine to
subdivide the simulated volume into a set of disjoint pieces that are
mapped onto the MPI ranks, allowing data replication to be kept to a
minimum.

The domain decomposition is redone for every time step whenever at least a
specified fraction of particles
$f_\text{domain} = N_\text{active} / N_\text{total}$ is active. In
order to allow for better simultaneous balancing of different cost factors
(which are the CPU-time needed to compute gravity and hydrodynamics,
possibly for several different time bins, and the memory load for
particles and cells), the computational domain is split into
$N_\text{domain}$ chunks,
\begin{align}
  N_\text{domain} = n_\text{domain} \, N_\text{MPI},
\end{align}
where $n_\text{domain}\ge 1$ is a free integer parameter, and
$N_\text{MPI}$ is the number of MPI ranks. By `oversampling' the number
of MPI ranks with $n_\text{domain}$ larger than unity, one gives the
code the ability to map several chunks onto the same MPI rank in order
to more effectively smooth out residual imbalances. This allows us, for
example, to combine on one MPI rank a chunk that has a lot of work
but few particles with one that has many particles but  requiring
little work, simultaneously balancing the memory and the work
load.

In practice, \textsc{Arepo} carries out the domain decomposition by
first constructing the highest levels of the oct-tree covering the full
simulation volume. The leaf nodes of this `top-level tree' are iteratively
replaced by refined leaf nodes if their 
load exceeds a target fraction
\begin{align}
  w_\text{opening} = \frac{w_\text{total}}{N_\text{domain} \, m_\text{domain}}
\end{align}
of the total load $w_\text{total}$.  Here the factor $m_\text{domain}$
is usually set to a value of a few $(\sim 2.5-4.0)$ to yield a
sufficiently fine discretization of the load at the level of the leaf
nodes, so that adjacent ones\footnote{Here adjacency is understood in
  a Peano-Hilbert sense.} can be combined into a final set of
$N_\text{domain}$ chunks that have, to a good accuracy, very similar
loads. These domain chunks are finally mapped to the $N_\text{MPI}$
processes, where for $n_\text{domain} > 1$ additional balancing
opportunities are realized. This secondary balancing step was not
realized in the \textsc{Gadget-2} code.

To estimate the work-related CPU-cost associated with particles and
cells in a leaf node, \textsc{Arepo} uses cost factors associated
with them. Specifically, to estimate the computational cost of the
hydrodynamics calculations associated with a domain patch, 
\begin{align}
c_\text{hydro} = \sum\limits_i b_i
\end{align}
is used, where $i$ runs over all particles located in the domain
piece, and the factor $b_i$ counts how often the particle
will be active in traversing the time step hierarchy before the  
next domain decomposition is expected to take place. For the
gravity calculations, a similar formula is used,
\begin{align}
c_\text{grav} = \sum\limits_i  g_i \, b_i ,
\end{align}
but here also a cost factor $g_i$ enters that measures the number of
particle-cell interactions that need to be evaluated for the particle
in the most recent gravity calculation. Depending on clustering state,
the values for $g_i$ can vary significantly, whereas a hydrodynamical
cell always has a very similar number of neighbors on
average. Finally, for the memory load, each collisionless particle and
hydrodynamical cell contribute equally in their respective category.
The parameters of the load-balancing algorithm are listed in
Table~\ref{TableLoadBalancing}.

\begin{table}
\caption{Code parameters for load balancing.}\label{TableLoadBalancing}
\centering
{
\begin{tabular}{|l|c|c|}
  \hline
  Description & Symbol & Code Parameter \\
  \hline \hline
   Domain decomposition frequency & $f_\text{domain}$ & \texttt{ActivePartFracForNewDomainDecomp} \\
  \hline
   Number of domains per MPI task & $n_\text{domain}$ & \texttt{MultipleDomains} \\
  \hline
   Depth parameter of the top-level tree & $m_\text{domain}$  & \texttt{TopNodeFacto}r \\
  \hline
\end{tabular}
}
\end{table}

\section{Other code features}
\label{sec:other}

\begin{table}
\caption{Code parameters for structure finding.}\label{TableSubfind}
\centering
{
\begin{tabular}{|l|c|c|}
  \hline
  Description & Symbol & Code Parameter \\
  \hline \hline
   FOF linking length & $l_\text{FOF}$ & \texttt{FOF\_LINKLENGTH} \\
  \hline
    Number of neighbors for subfind density estimate & $N_\text{ngb, sub}$ & \texttt{DesLinkNgb} \\
  \hline
    Tree opening criterion in subfind & $\vartheta_\text{sub}$ & \texttt{ErrTolThetaSubfind} \\
  \hline
\end{tabular}
}
\end{table}

Besides the core functionality to advance a simulation in time,
\textsc{Arepo} has a number of additional features that influence
practical aspects of how the code can be used in specific situations.
These are described in the following.

\subsection{On-the-fly structure and substructure finding}

\textsc{Arepo} includes a friends-of-friends (FOF) and a substructure
identification algorithm that can be used both on-the-fly or in
post-processing.  The algorithms are originally described in
\citet{Springel+2001}, and can be optionally called before each
snapshot dump, or in other intervals.

In brief, group and subhalo identification works in the following way.
First, an FOF algorithm is applied to the particle
distribution to define groups as equivalence classes, where any pair
of particles is in the same group if they are closer to each other
than $l_\text{FOF}$ times the mean inter-particle separation. Next,
these groups are subjected to substructure identification with the
\textsc{Subfind} algorithm \citep{Springel+2001}.

To this end, the local density is estimated in an SPH-like approach at
the position of all member particles, by using a smoothing length that
encloses $N_\text{ngb, sub}$ (weighted) nearest neighbors. Then, an
excursion set algorithm is used to identify locally overdense regions
as substructure candidates.  Each candidate is then treated with a
gravitational unbinding procedure that iteratively excludes all
particles with positive total energy from the member list of the
substructure candidate. The potential binding energy is calculated
using a tree algorithm analogous to the one used for the gravitational
force calculation, with opening criterion $\vartheta_\text{sub}$. If a
gravitationally bound set of particles survives that contains at least
a minimum number of particles, the substructure is retained in the list
of final subhalos. The algorithm can detect nested sets of subhalos,
but each particle/cell is counted only toward the mass of one
substructure. Groups and subhalos are allowed to be much larger than
the ones that fit into the memory of a single MPI rank.

The particles and cells of a snapshot dump are automatically stored on-disk 
such that particles belonging to individual (FOF) groups are
grouped in the output as one block, in order of descending group size.
Within a given group, the particles are further sorted according to
subhalo membership, again in descending order of subhalo
size. Finally, within a subhalo, the particles are sorted by their
binding energy, with the particle with the lowest total energy coming
first. This allows efficient random access to the particle data
belonging to groups or subhalos even of very large simulation
outputs. More details on the structure of the simulation output when
using \textsc{Subfind} can be found in the data release papers of the
Illustris and IllustrisTNG projects \citep{Nelson+2015, Nelson+2018b}.
The parameters for the structure and substructure finding algorithms
are listed in Table~\ref{TableSubfind}.

\subsection{Initial conditions conversion}

Due to its ancestry in the SPH code \textsc{Gadget}, \textsc{Arepo}
includes an option to convert SPH initial conditions to ones suitable
for grid-based calculations. To this end, first an oct-tree structure
is built for the computational domain using the SPH particles,
augmented with a coarse grid of massless background particles to fill
the computational volume of an enclosing box.  The leaf nodes of this
tree form a tessellation of space, which is now used for distributing
the SPH quantities in a conservative fashion to the leaf nodes using
the SPH kernel. Finally, a mesh-generating point is placed at the
cell-centers of the leaf nodes, inheriting the conserved fluid
quantities that were scattered by the SPH particles to these points.  In
this way, grid-based initial conditions for \textsc{Arepo} are
generated that mimic the SPH distribution as closely as possible, but
without having a highly irregular initial mesh, which would result if
one uses the SPH particles directly (which is also possible).

One of the provided examples included with \textsc{Arepo}, a merger of
two galaxies, demonstrates how this is done in practice. We want to
emphasize, however, that the initial conditions created in this way
may still suffer from Poisson-like discreteness noise of SPH
snapshots, and directly creating new, grid-based initial conditions
for a known smooth hydrodynamic field will result in higher-quality
initial conditions. Therefore, whenever possible, we recommend to
rather modify the initial conditions generating code to support
grid-based setups over converting SPH initial conditions.

\subsection{Input and output}

The standard input and output of simulation time slices is done via
snapshot files, usually employing the HDF5
format\footnote{https://www.hdfgroup.org/solutions/hdf5/}. Older
binary formats identical to the ones used in \textsc{Gadget-2}
(referred to as format 1 and 2) are also supported. Both initial
conditions and snapshot outputs can be distributed over an arbitrary
number of files\footnote{For outputs, the number is limited to be at
  most equal to the number of employed MPI ranks.}.  The number of
desired files is recognized automatically for inputs, while for
outputs it is specified as a parameter, which can be chosen both for
convenience (to limit the maximum size of individual files, which
helps in transferring/archiving large snapshots) and for increasing the
achieved I/O bandwidth, depending on the system.  This is because the
files belonging to an input/output set can be read/written in parallel,
with each file being assigned to a specific MPI task, normally chosen
automatically to be maximally spread over the used compute nodes as
well. These tasks then handle the I/O for a group of MPI ranks and
collect/send the corresponding data over the MPI communication
fabric. In this way, highly efficient parallel I/O can be achieved
that makes full use of the maximum I/O bandwidth available on a
particular parallel file system. In order to not overload a given
system (and to retain responsiveness for other users), one can also
specify a maximum number of files written/read in parallel at any
given time.

Note that in this approach, each individual file is written with
ordinary POSIX semantics, i.e., \textsc{Arepo} does neither make use of
the parallel I/O functionality of HDF5 nor employs MPI-IO
functions. The explicit parallel I/O approach we prefer minimizes
side effects of these complex libraries, such as their internal memory
allocation calls, which can become prohibitive when physical memory is
extremely scarce, a situation that (unfortunately) often occurs in
practice for large production runs.

\subsubsection{Binary dumps for restarting the code}

Since the run time of simulations is often longer than the
wall-clock time limits of queuing systems on compute clusters, a single
submission of \textsc{Arepo} may not be enough to finish a
simulation. The code therefore includes a check-pointing functionality,
where so-called restart files are written by each MPI rank separately,
allowing a seamless continuation of a run in a subsequent
submission. Particular care has been taken to ensure that resuming
from such a restart dump yields binary-identical results compared to
continuously executing the code without a break (changing the number
of MPI ranks upon restarting is thus not possible).

The restart files are essentially binary dumps of all necessary
simulation variables in their full precision. They are written
automatically at the end of a specified run time, and also in regular
time intervals to guard against losing time due to a node or code
failure. Copies of the two most recent restart files
are kept automatically, and replaced in a cyclic fashion.

For large, long-running simulations (such as IllustrisTNG), the
cumulative data volume written to disk can become extremely large, to
the point that potential file system errors can be no longer ignored,
despite them being exceedingly rare. To this end, \textsc{Arepo} makes
special efforts to detect and avoid these for the critical restart
files.  After writing a set of restart files, md5-checksums of the
data in memory are compared with md5-checksum of the file on-disk when
read-in by an MPI rank on a \textit{different} compute node. We found
this to be a reliable way to detect corrupted files, in which case the
code will terminate to ensure that the user is made aware of the file
system problems, and in order to leave one of the two sets of restart
files on-disk intact.

\subsection{Memory management}

Computational power tends to increases faster than the amount of
physical memory available on modern computers. Cosmological
simulations are therefore often memory bound and not necessarily CPU-time
bound. It is therefore important to make optimal use of the available
physical memory. This, in particular, calls for complete control over
the amount of memory used by the simulation code at any given time,
the safe prevention of any memory leak, and the avoidance of
fragmentation of the memory heap\footnote{Fragmentation can especially
  become an issue on compute nodes with a simplified virtual memory
  allocation system, such as the Bluegene.}.  In \textsc{Arepo}, this
is achieved by an internal memory manager that, upon start-up, allocates a
single block of memory from the system for each MPI task, and then
handles all of the dynamic memory allocation requests internally using
this storage space.  In this way, the maximum amount of physical
memory available to application codes can robustly be used by
\textsc{Arepo} (modulo the sometimes uncertain requirements of the MPI
library, or of background system processes related to, e.g., the Lustre
parallel file system). In case one of the MPI ranks running out of
memory, \textsc{Arepo} itself terminates and provides a verbose
overview of the allocated memory, including information about which
line of the code allocated each array or variable.

In order to avoid fragmentation, the internal memory manager normally
enforces that allocated memory blocks are freed in inverse order of
the allocation, such that it is used like a stack. However, also
movable memory blocks can also be used in special situations where this
leads to simpler or more efficient code, so that the memory manager
does not impose a serious restriction on the programming model.

\subsection{Usage of libraries}
\label{sec:usage_of_libraries}

The usage of libraries in codes can reduce the development time and
improve the reliability of individual code parts. An overly excessive
use of (exotic) libraries, however, can also cause problems for
building and maintaining the code, and it introduces a source of
potential errors and performance problems outside of the direct control of
the simulation code developers. In \textsc{Arepo}, the general
philosophy is, for this reason, to limit the number of libraries used to
an essential minimum. In practice, this means that \textsc{Arepo}
makes use only of common and well-tested libraries that are usually
already present on compute clusters and supercomputers. These are MPI
for parallelism, HDF5 for I/O, GSL for a few elementary numerical
integrations and pseudo random number generation, FFTW for FFTs, 
and finally the GMP library for arbitrary precision
integer arithmetic. The latter is employed for robustly resolving
degeneracies in the mesh construction.

\textsc{Arepo} is written in C, using the C11 standard. Consequently,
it can be compiled with a large number of different compilers. We
recommend the free and widely used GNU compiler for its excellent
robustness and speed. For \textsc{Arepo}, our experience is that
commercial compilers only sometimes result in slightly better
performance of the code on some machines, but are also more prone to
producing incorrect code optimizations when aggressive optimization
levels are used. The code employs the \textsc{MPI-2} standard for
distributed-memory parallelism, and exercises only a comparatively
small subset of all possible MPI functions.  In the public version of
\textsc{Arepo}, only communication patterns that, in our experience, are
very reliable on most machines are used.

\section{Examples}
\label{sec:testcases}

\begin{table}[h!]
\caption{Implemented simulation examples.}
\label{table:tests}
\centering
\begin{tabular}{|l|c|c|}
\hline
\textbf{Name of Example} & \textbf{Description} & \textbf{Verification} \\
\hline \hline
wave\_1d & one-dimensional wave propagation & analytical solution \\
shocktube\_1d & one-dimensional Riemann problem & exact solution \\
interacting\_blastwaves\_1d & one-dimensional interacting shocks & high-resolution simulation \\
mhd\_shocktube\_1d & one-dimensional MHD Riemann problem & {exact solution} \\
polytrope\_1d\_spherical & hydrostatic $n=1$ polytrope & initial conditions \\
noh\_2d & converging flow in 2D & analytic solution \\
gresho\_2d & stationary vortex & initial conditions \\
yee\_2d & stationary vortex & initial conditions \\
current\_sheet\_2d & magnetic field reversal & magnetic energy \\
noh\_3d & converging flow in 3D & analytic solution \\
cosmo\_box\_gravity\_only\_3d & gravity-only cosmological structure formation & halo mass (reference run) \\
cosmo\_box\_star\_formation\_3d & star formation in cosmological volume & stellar mass density (reference run)\\
cosmo\_zoom\_gravity\_only\_3d & gravity-only cosmological zoom simulation & subhalo mass (reference run)\\
galaxy\_merger\_star\_formation\_3d & merger of two isolated disk galaxies & stellar mass (reference run) \\
isolated\_galaxy\_collisionless\_3d & isolated disk galaxy (stars and dark matter) & disk scale height \\
\hline
\end{tabular}
\end{table}

\textsc{Arepo} includes a number of small example simulations that
are also used for code verification. They are implemented in such a
way that they can be (and are) used for regular automated code
regression tests to ensure that the functionality and correctness of
the code under certain conditions, and especially after changes have
been made, is not broken. The default set of tests is listed in
Table~\ref{table:tests}.

\subsection{Wave propagation}

\begin{figure}
  \includegraphics{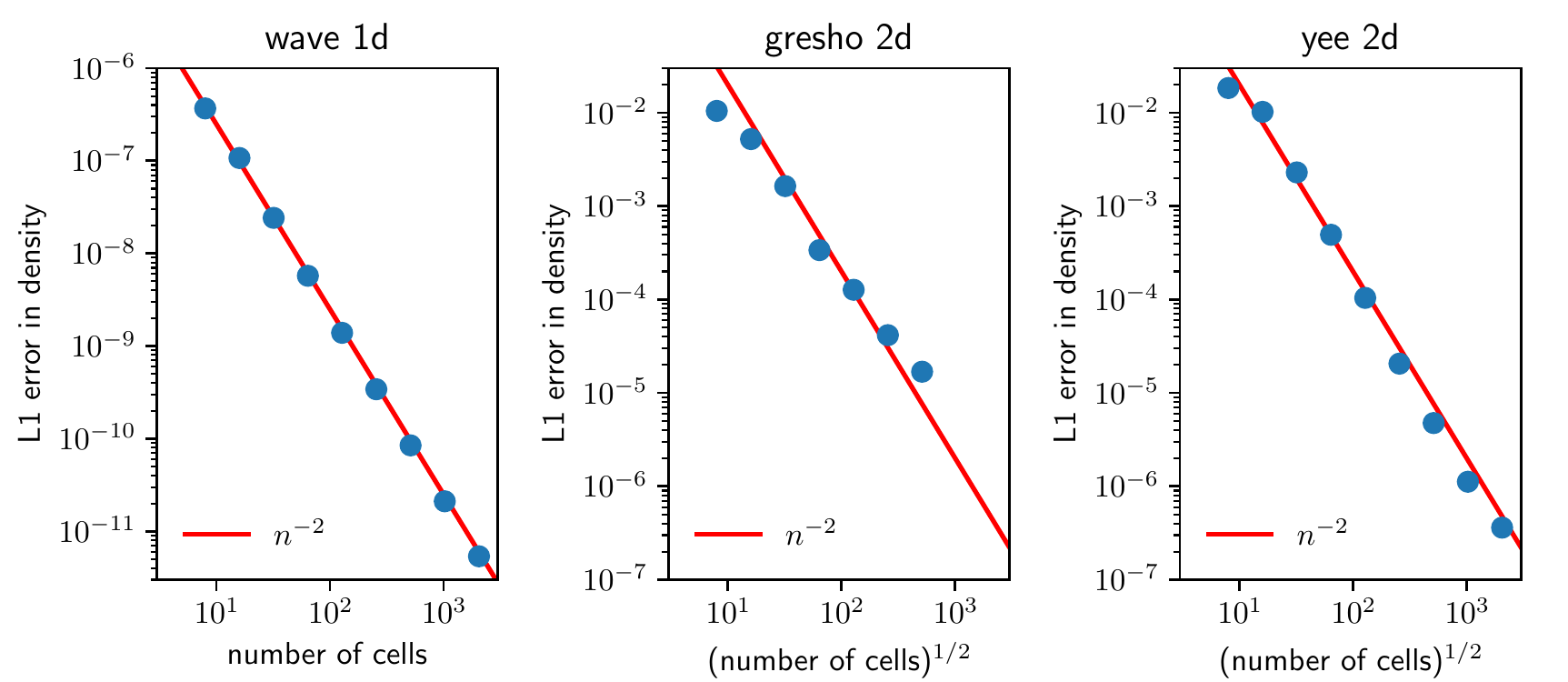}
  \caption{{Convergence of the density field for the wave problem (left), the 
  Gresho vortex (center), and the Yee vortex (right). Wave propagation and
  the Yee vortex show second-order convergence, the Gresho vortex shows 
  lower-than-second-order convergence, in qualitative agreement with previous 
  findings \citep{Springel2010}. This is likely caused by the non-smooth
  initial conditions of this problem.}  }
    \label{fig:convergence}
\end{figure}

One of the most elementary tests for a hydrodynamic code is the
propagation of a small amplitude sound wave in one dimension. We use a
1D periodic domain of length $L=1$, filled with a fluid of density
$\rho_0 = 1$, velocity $v_0=0$, pressure $p_0=3/5$, and an adiabatic
index of $\gamma=5/3$. We then introduce a small sinusoidal adiabatic
perturbation of amplitude $\delta=10^{-6}$ and wavenumber $k=2\pi/L$,
such that density, velocity, and specific internal energy are given by
\begin{align}
  \rho(x) &= \rho_0 \left[ 1 + \delta \, \sin\left( k x \right) \right], \\
  v(x) &= 0, \\
  u(x) &= u_0 \left[ 1 + \delta \, \sin\left( k x \right) \right]^{\gamma - 1},
\end{align}
where $x$ is the spatial coordinate. We {use $C_\text{CFL} = 0.3$ and} 
simulate the problem until $t=1$, i.e., the time the perturbation takes to 
propagate once through the computational domain, and compare the result 
with the initial conditions. Since this is a smooth 
problem, it can also be used to test the convergence order of the 
hydrodynamics scheme, which, in our case, is particularly sensitive to
the gradient estimates {and shown in Figure~\ref{fig:convergence}, 
left panel.}

\subsection{Riemann problem}

\begin{figure}
  \includegraphics{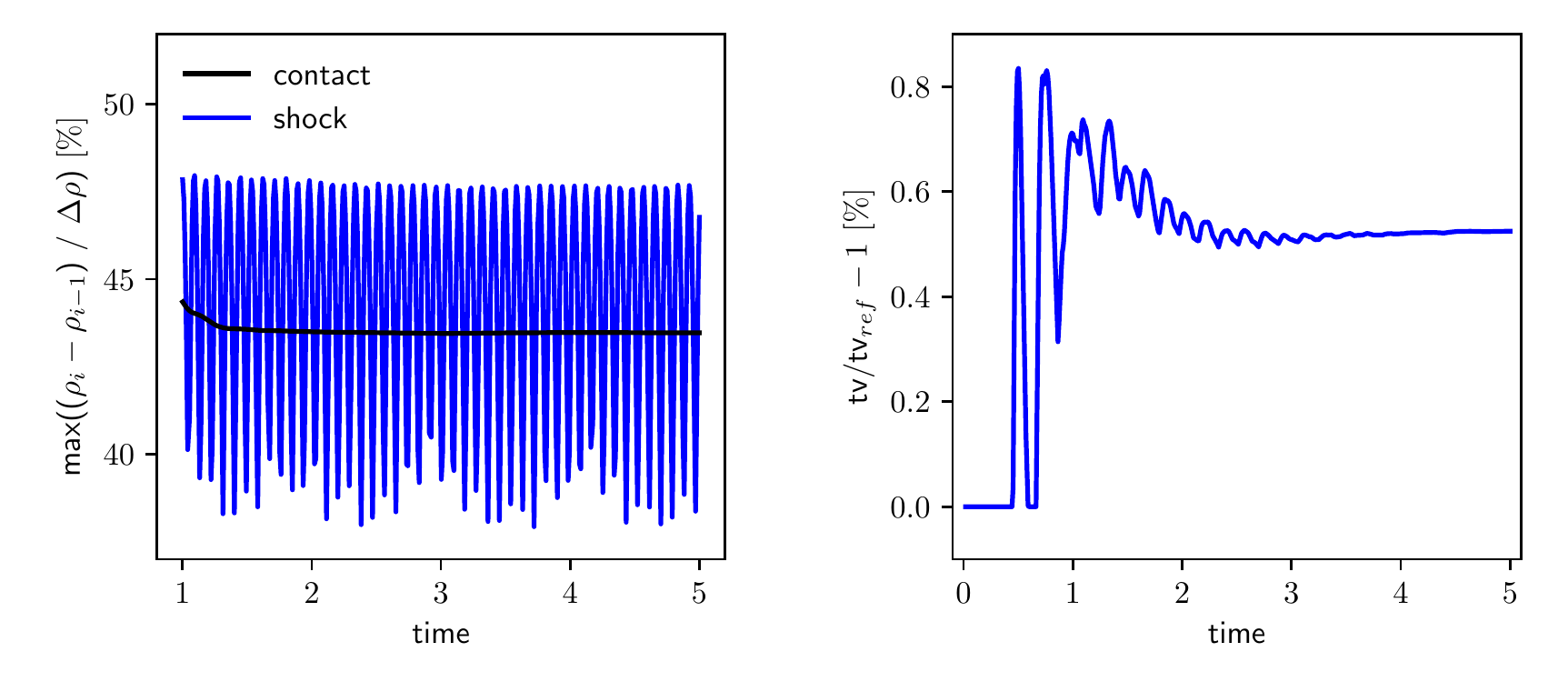}
  \caption{{Fraction of the discontinuity resolved by a single cell (left) for 
  the 1d shock tube test. The oscillation in the shock is a result of cells moving 
  through the shock interface, while the constant value at the contact discontinuity 
  is due to cells moving with the flow. The flux limiters cause the total variation to 
  increase by less than $1 \%$ (right), effectively avoiding spurious oscillations at
  discontinuities.}  }
    \label{fig:shocks}
\end{figure}

Another basic test is the calculation of a one-dimensional Riemann
problem, i.e., a domain split into a left-hand initial state and a
right-hand state. In our particular example, we set up a density
and pressure jump with no initial velocity on either side of the
initial separation at $x=10$, using a box with domain length $L=20$
and reflective boundary conditions. We use a moving mesh with $128$
initially equally spaced cells, without refinement
or de-refinement for this test. The density $\rho$, velocity $v$, and
pressure $p$ to the left and right of the discontinuity are given by
\begin{align}
 \rho_L = 1.0, \qquad & \rho_R = 0.125, \\
 v_L = 0, \qquad & v_R = 0, \\
 p_L = 1.0, \qquad & p_R = 0.1,
\end{align}
respectively, and an adiabatic index of $ \gamma = 1.4$ is assumed 
\citep[see][]{Sod1978}{, and $C_\text{CFL} = 0.3$.}
These initial conditions lead to a shock traveling to the right, a
rarefaction wave traveling to the left, and a moving contact
discontinuity in the center. In this test, we check not only against
the iteratively computed solution, but also monitor the total
variation of the solution, i.e., the sum of absolute differences in the
hydrodynamical quantities of neighboring cells, to ensure that the
slope limiters prevent oscillatory instabilities from
growing. Similarly, we analyze the physical discontinuities and check
that the jumps in density, velocity, and pressure, are resolved with
at most a few cells across (four has proven to be a suitable value for demanding
that the 5th and 95th percentiles of the jump are caught). This sets
strong limits on the allowed numerical diffusivity of the
hydrodynamical scheme. {Figure~\ref{fig:shocks} shows the 
maximum change in density from a single cell to the next relative to 
the analytic discontinuity (left) as well as the total variation of the density
relative to the analytic solution (right) as a function of time.}

\subsection{Interacting blastwaves}

\begin{figure}
  \includegraphics{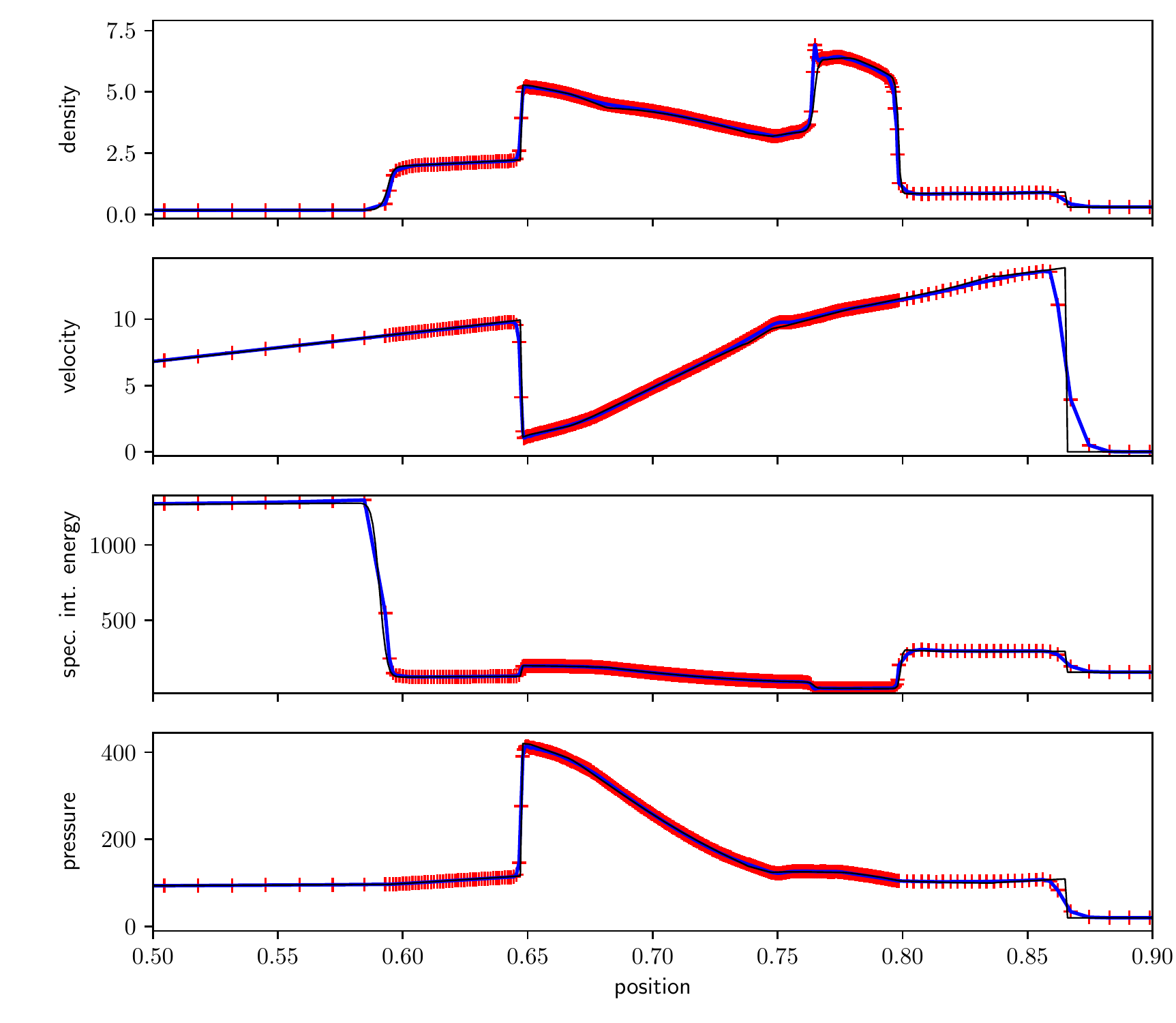}
  \caption{{Hydrodynamic quantities of an interacting blast wave problem at 
  $t=0.038$. The black line shows the high-resolution reference solution, and the 
  blue line shows the moving-mesh solution with the red crosses denoting the cell
  positions.}}
  \label{fig:interacting_shocktube_1d}
\end{figure}

A slightly more complex, and also numerically more challenging, test is
a problem of interacting shocks as described, e.g., in
\citet{Springel2010}. We set up a 1D domain with length $L=1$ and
reflective boundary conditions. Initially, two Riemann problems are
set up at $x=0.1$ and $x=0.9$, leaving three hydrodynamic states in
the initial conditions. We choose as density, velocity, and pressure,
for the left, center, and right states:
\begin{align}
  \rho_L &= \rho_C = \rho_R = 1, \\
  v_L &= v_C = v_R = 0, \\
  p_L &= 1000,\qquad p_C = 0.1,\qquad p_R = 100,
\end{align}
respectively. The adiabatic index of the gas is $\gamma = 1.4$. We
allow the initially equally spaced cells to move with the flow, but do
not include refinement or de-refinement. As in \citet{Springel2010},
the problem is evolved until $t=0.038$ 
{using $C_\text{CFL} = 0.3$}, and at this time is compared
to a very high-resolution, static grid solution of the same initial
conditions (see Figure~\ref{fig:interacting_shocktube_1d}).

\subsection{MHD shocktube}

\begin{figure}
  \includegraphics{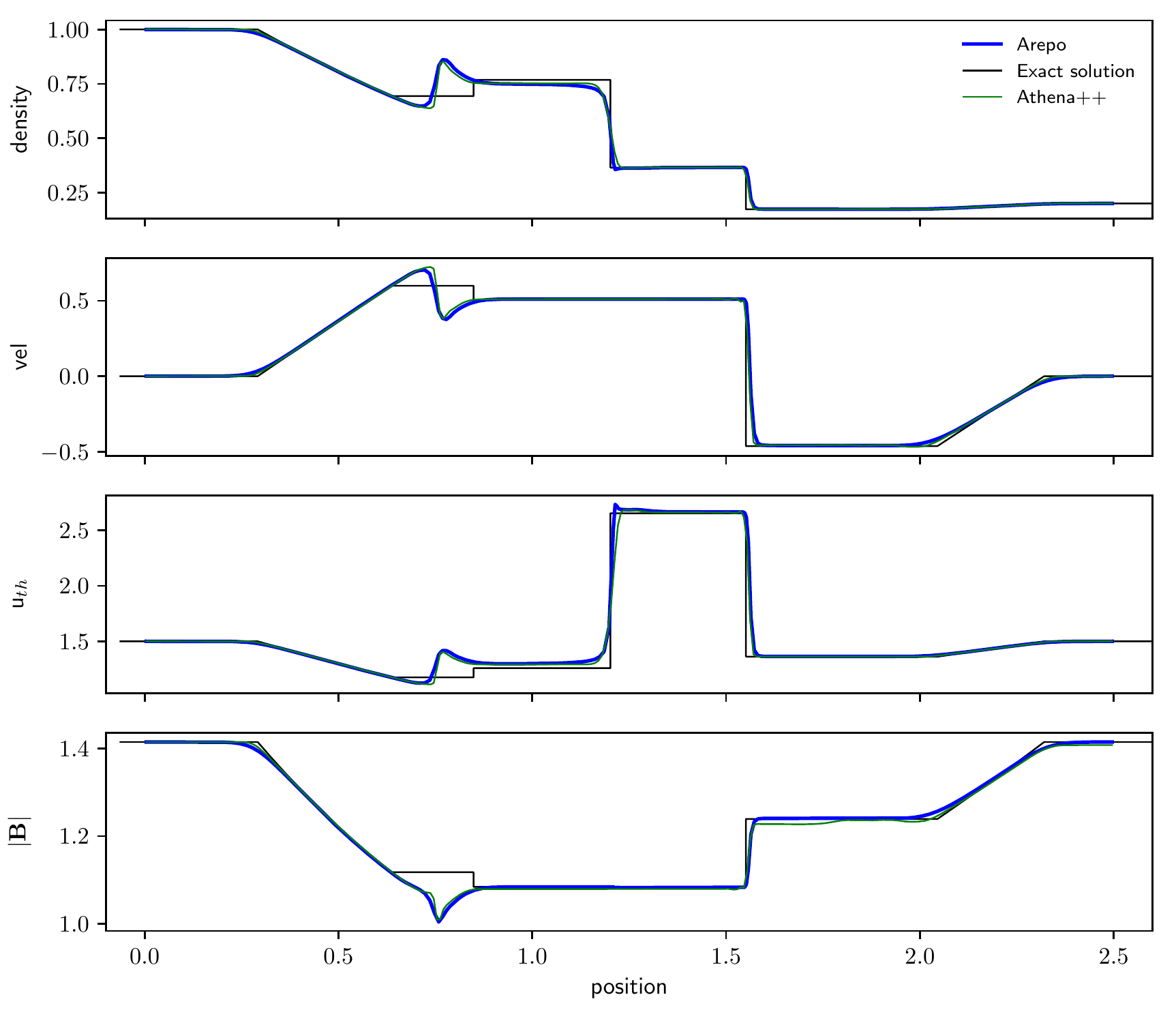}
  \caption{{Magnetohydrodynamic properties of an MHD shocktube problem
    at time $t=0.4$. The black line shows the exact solution, while the blue and green 
    lines show solutions computed with \textsc{Arepo} and \textsc{Athena++}, respectively, using
    300 cells.} }
    \label{fig:mhd_shocktubne_1d}
\end{figure}

\begin{figure}
  \includegraphics{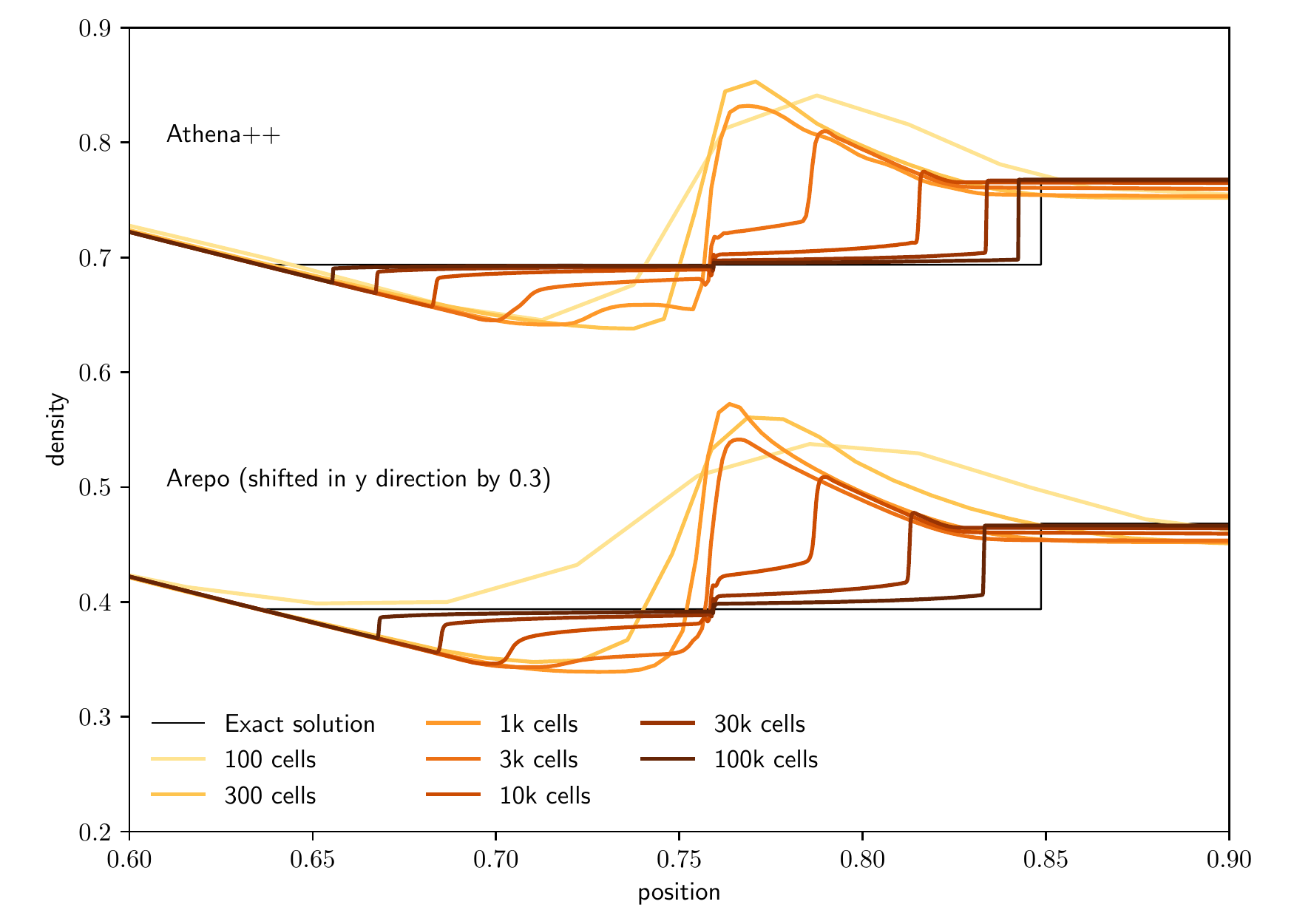}
  \caption{{Density of an MHD shocktube problem at time $t=0.4$ focused on positions
    between $0.6$ and $0.9$ for different resolutions. 
    Both codes, \textsc{Arepo} and \textsc{Athena++} converge toward the exact 
    solution, but only at very high resolution. The convergence in the \textsc{Arepo} 
    case is slightly slower, requiring a factor 
    of three more cells to produce a comparable result.}}
    \label{fig:mhd_shocktubne_1d_convergence}
\end{figure}

To test the MHD Riemann solver in \textsc{Arepo}, we use a magnetic
Riemann problem as described in \citet{Torrilhon2004} and \citet{Fromang+2006}. We employ
a 1D domain with length $L=2.5$ and reflective boundary
conditions. We set up a discontinuity at $x=1$ with density,
velocity, pressure, and the three components of the magnetic field to
the left and right, given as
\begin{align}
  \rho_L = 1.0, \qquad & \rho_R = 0.2, \\
  v_L = 0.0, \qquad & v_R = 0.0, \\
  p_L = 1.0, \qquad & p_R = 0.2, \\
  bx_L = 1, \qquad & bx_R = 1, \\
  by_L = 1, \qquad & by_R = \cos(\alpha), \\
  bz_L = 0, \qquad & bz_R = \sin(\alpha),
\end{align}
respectively\footnote{{Note that the magnetic field is given here in
Heaviside-Lorentz units, while in the initial conditions, Gaussian units are
used.}}. In this example, we use $\alpha = 3$, and the
adiabatic index of the thermal fluid is $\gamma=5/3$. We allow the
initially uniform mesh to move with the fluid flow, but refrain from
using refinement or de-refinement. {The employed Courant factor is
 $C_\text{CFL} = 0.3$.} The solution is verified at time
$t=0.4$ against {the exact solution} (Figure~\ref{fig:mhd_shocktubne_1d}).
{We additionally perform the same simulation using the \textsc{Athena++} 
code \citep[][Stone et al. 2020, in preparation]{White+2016, Felker+2018}\footnote{The setups for the \textsc{Athena++} runs are available under {https://github.com/rainerweinberger/athena-public-version.git}, input files \texttt{athinput.tb} and \texttt{athinput.current\_sheet}.}.
Similarly to \citet{Fromang+2006}, the low-resolution version shows significant errors, 
and only when increasing the resolution substantially can
the exact solution be recovered (Figure~\ref{fig:mhd_shocktubne_1d_convergence}). 
We find the same behavior both for \textsc{Arepo}
and \textsc{Athena++}; however, roughly a factor of three higher resolution is required in \textsc{Arepo} 
compared to \textsc{Athena++} to produce a similar degree of accuracy in this problem.}

\subsection{Static polytrope}

\begin{figure}
  \includegraphics{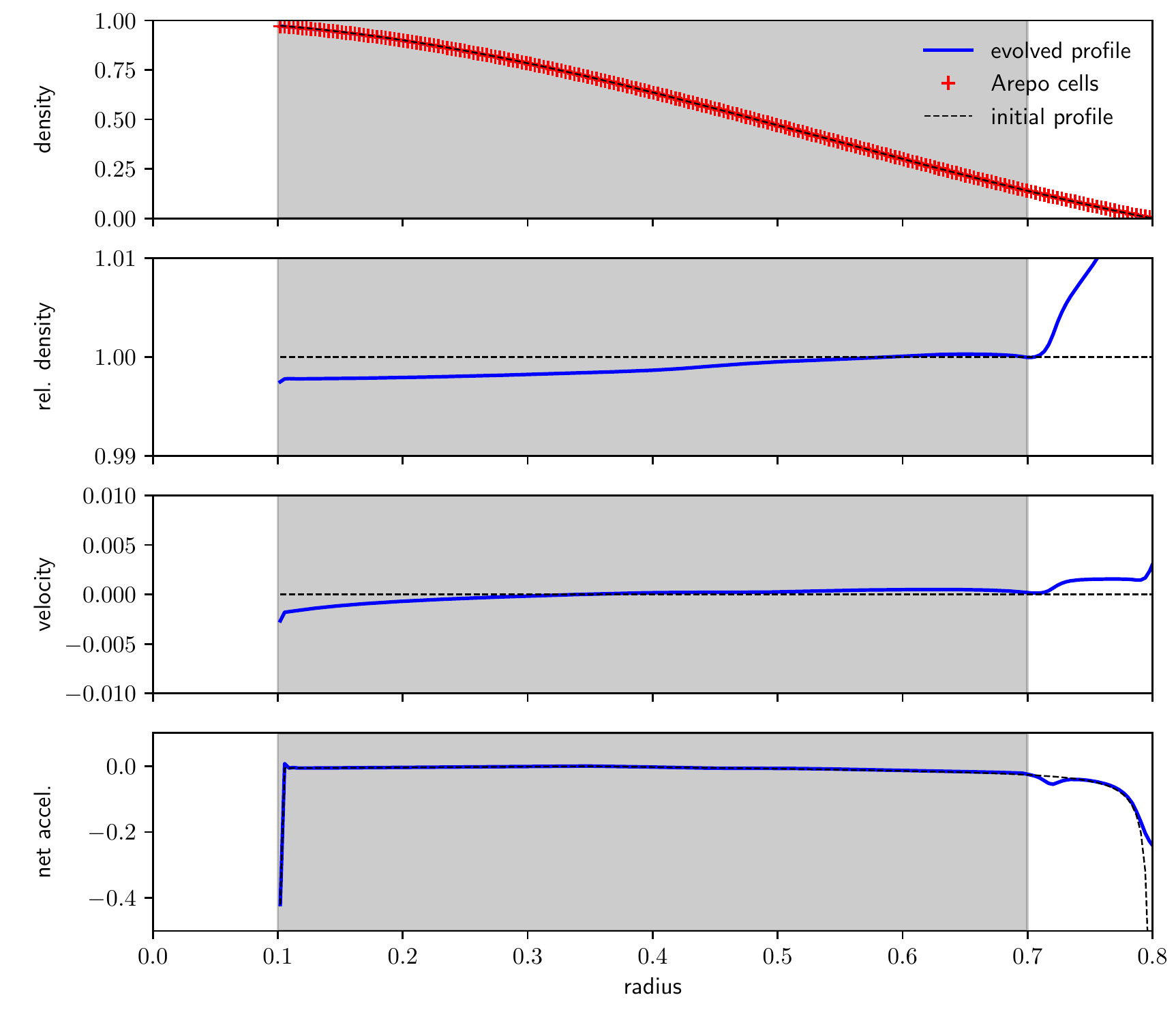}
  \caption{Density profile, relative density deviation from initial conditions, radial velocity, and net acceleration as a function of radius for a 1D spherically symmetric $n=1$ polytrope setup.}
  \label{fig:polytrope_1d_spherical}
\end{figure}

To test the 1D spherically symmetric mode of \textsc{Arepo}, we set up
a known hydrostatic solution of the $n=1$ polytrope. This setup uses a
reflective inner and an open outer boundary condition with a
computational domain from $r=0.1$ to $r=1$.  For the $n=1$ polytrope
(with $\gamma = 2$), a static solution
\begin{align}
 \theta(\xi) = \frac{\sin(\xi)}{\xi}
\end{align}
exists, with $\xi = {r}/{r_\text{scale}}$ being the dimensionless
radius and $\rho = \rho_0\,\theta$. In this particular setup, we use
$r_\text{scale}=0.8/\pi$, $\rho_0=1$, and $G=1$, in simulation units.
The pressure is given at each radius by $p = K \rho^\gamma$ with
$K=2 \pi G r_\text{scale}^2$, and the initial velocity is $v=0$
everywhere.  For gravity, besides the gravitational force of
the enclosed gas mass, we include a static core with radius $r=0.1$
and mass $m=4.18\times 10^{-3}$. {The employed Courant factor is
 $C_\text{CFL} = 0.3$, and} the mesh is kept static in this
problem.  Since this problem is set up in hydrostatic equilibrium,
verification is done by comparing the result at $t=1$ to the initial
conditions (see Figure~\ref{fig:polytrope_1d_spherical}).

\subsection{Magnetic current-sheet problem}
\label{sec:current_sheet_2d}

\begin{figure}
  \includegraphics{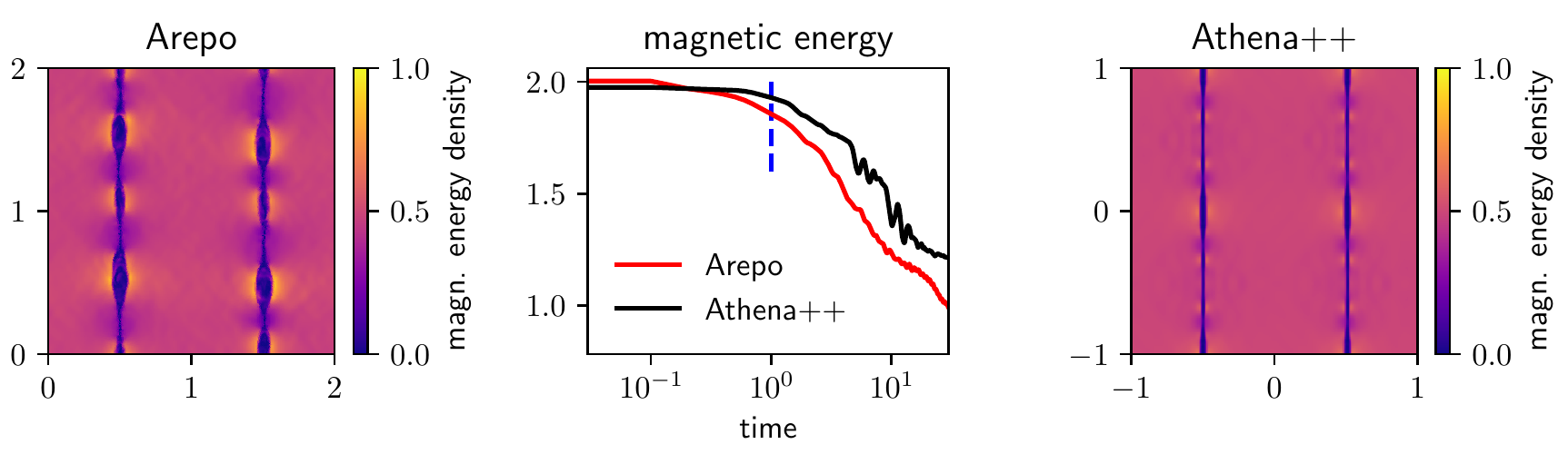}
  \caption{{Left and right panels: magnetic energy density of the current-sheet 
  problem at time $t=1$ computed with \textsc{Arepo} and \textsc{Athena++}
  , respectively, each using $256^2$ cells. 
  The \textsc{Arepo} result shows a significantly growing instability driven by the larger
  numerical reconnection, while the \textsc{Athena++} simulation remains relatively 
  static to this point. Eventually, numerical reconnection will, in both cases, lead 
  to a growth of the instability as can be seen in the time evolution of the 
  magnetic energy in the simulation (central panel).}}
  \label{fig:current_sheet_2d}
\end{figure}

Another test of the MHD scheme, in particular for the
numerical reconnection rates, is the so-called current-sheet problem,
following \citet{Gardiner+2005}.  We set up a two-dimensional domain
with side length $L=2$ and periodic boundary conditions. The
magnetohydrodynamic quantities are
\begin{align}
  \rho &= 1, \\
  v_x &= 0.1 \sin(\pi y), \\
  v_y &= 0, \\
  p &= 0.1, \\
  b_x &= 0, \\
  b_y &=
  \begin{cases}
  1 \qquad \text{if } x < 0.5 \text{ or } x>1.5, \\
  -1 \qquad \text{else},
  \end{cases}
\end{align}
and an adiabatic index $\gamma = 5/3$ is used\footnote{{Note 
that the magnetic field here is given in Heaviside-Lorenz units,
while the \textsc{Arepo} input is given in Gauss units.}}. The cells are allowed
to move, but refinement and de-refinement are disabled, {and 
 $C_\text{CFL} = 0.3$.  To verify the solution, 
 we check the total magnetic energy in the system. For comparison, 
 we ran the same problem using the \textsc{Athena++} code. 
Figure~\ref{fig:current_sheet_2d} shows slices
 of the magnetic energy density at $t=1$ and the time evolution of the 
 total magnetic energy in the simulation until $t=30$, showing a higher 
 numerical reconnection rate in the \textsc{Arepo} simulation}.

\subsection{Noh shock problem}

\begin{figure}
  \includegraphics{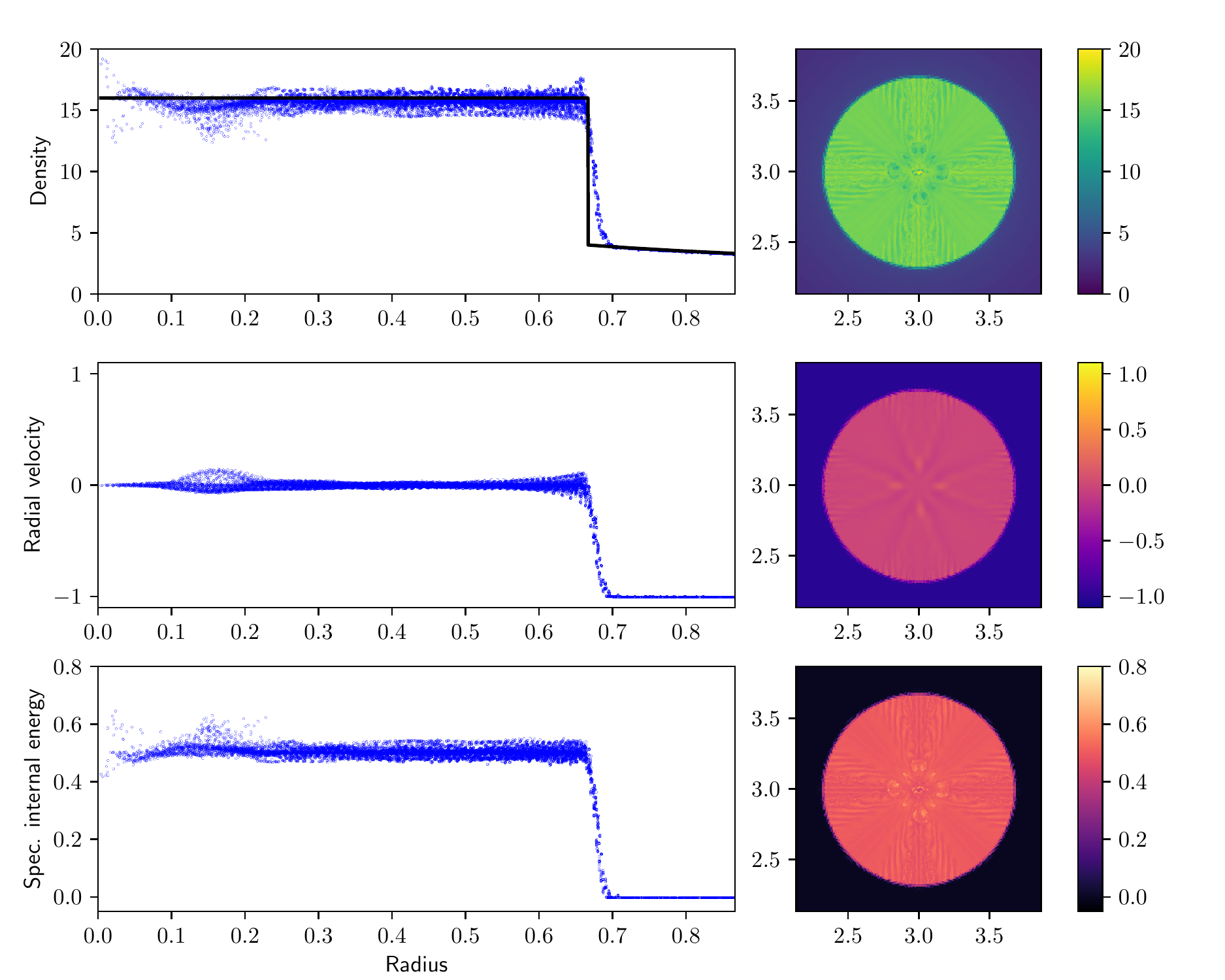}
  \caption{{Two-dimensional Noh problem at time $t=2$ using $150^2$ initially uniformly spaced grid cells.}}
  \label{fig:noh_2d}
\end{figure}

As a test for the Riemann solver in two and three dimensions, we use
Noh's shock problem \citep{Noh1987} in a computational domain of side length $L=6$,
with open boundary conditions. Initial density, velocity, and pressure
are given by
\begin{align}
  \rho &= 1, \\
  \vec{v} &= - \hat{r}, \\
  p &= 10^{-4},
\end{align}
where $\hat{r}$ is the unit vector of the position relative to the
domain center, i.e., we set up a constant velocity inflow to the
center.  The adiabatic index of the gas is $\gamma = 5/3$, {and the 
Courant factor $C_\text{CFL}=0.3$}. The cells
are allowed to move, without considering refinement or
de-refinement. The inflowing material causes a strong shock to develop
in the center, traveling outwards. The (analytically derivable)
post-shock solution as well as the position of the shock are then
checked in the  verification {(see Figure~\ref{fig:noh_2d})}.

\subsection{Gresho vortex}

In two dimensions, it is interesting to test the hydrodynamical scheme
via a stationary vortex problem, i.e., a problem where the centrifugal
forces are balanced by a pressure gradient. One such example is the
Gresho vortex problem \citep{Gresho+1990}. We set up a two-dimensional box with side
length $L=1$ and periodic boundary conditions. For the density,
azimuthal velocity, and pressure, we choose
\begin{align}
  \rho &= 1, \\
  v_\varphi &=
  \begin{cases}
  5\, r\qquad &\text{for }r<0.2 ,\\
  2 - 5\, r \qquad &\text{for }0.2\leq r < 0.4, \\
  0 \qquad &\text{for } r \geq 0.4, \\
 \end{cases} \\
  p &=
  \begin{cases}
  5 + 12.5 \, r^2,\qquad &\text{for }r<0.2 ,\\
  9 + 12.5 \, r^2 - 20 \, r + 4.0 \, \ln(r / 0.2) \qquad &\text{for }0.2\leq r < 0.4, \\
  3 + 4  \ln(2) \qquad &\text{for } r \geq 0.4.
  \end{cases}
\end{align}
The initial radial velocity is zero. We {use $C_\text{CFL} = 0.3$ and} 
let the problem evolve with a moving mesh until $t=3$, and verify the result 
by comparing to these initial conditions.  {We study the convergence using 
the density field in Figure~\ref{fig:convergence}, central panel.}

\subsection{Yee vortex}

The Yee vortex problem \citep{Yee+2000} is similar in nature to the Gresho vortex,
however, has the advantage of having a smooth solution. We follow the
setup of \citet{Pakmor+2016} and use a two-dimensional computational
domain with side length $L=10$ and periodic boundary conditions. Adopting
a temperature at infinity of $T_\text{inf} = 1.0$, $\beta=5.0$, and an
adiabatic index $\gamma = 1.4$, the temperature profile is given by
\begin{align}
T(r) = T_\text{inf} - \frac{ (\gamma - 1) \beta^2 }{8 \pi^2 \gamma} \exp(1 - r^2),
\end{align}
and the density $\rho$, azimuthal velocity $v_\varphi$ and specific
internal energy $u$ as a function of radius $r$ are given by
\begin{align}
  \rho(r) = T(r)^\frac{1}{\gamma - 1}, \\
  v_\varphi(r) = \frac{r \beta}{2 \pi} \exp\left( \frac{1 - r^2}{2}\right) ,\\
  u(r) = \frac{T(r)}{\gamma - 1}.
\end{align}
The radial velocity is zero.  We  {employ $C_\text{CFL} = 0.3$ and} 
evolve this setup with a moving mesh until $t=10$ and again verify the 
result by comparing with the initial conditions. 
{We study the convergence using the density field in 
Figure~\ref{fig:convergence}, right panel.}

\subsection{Cosmological simulation with gravity only}

\begin{figure}
  \includegraphics{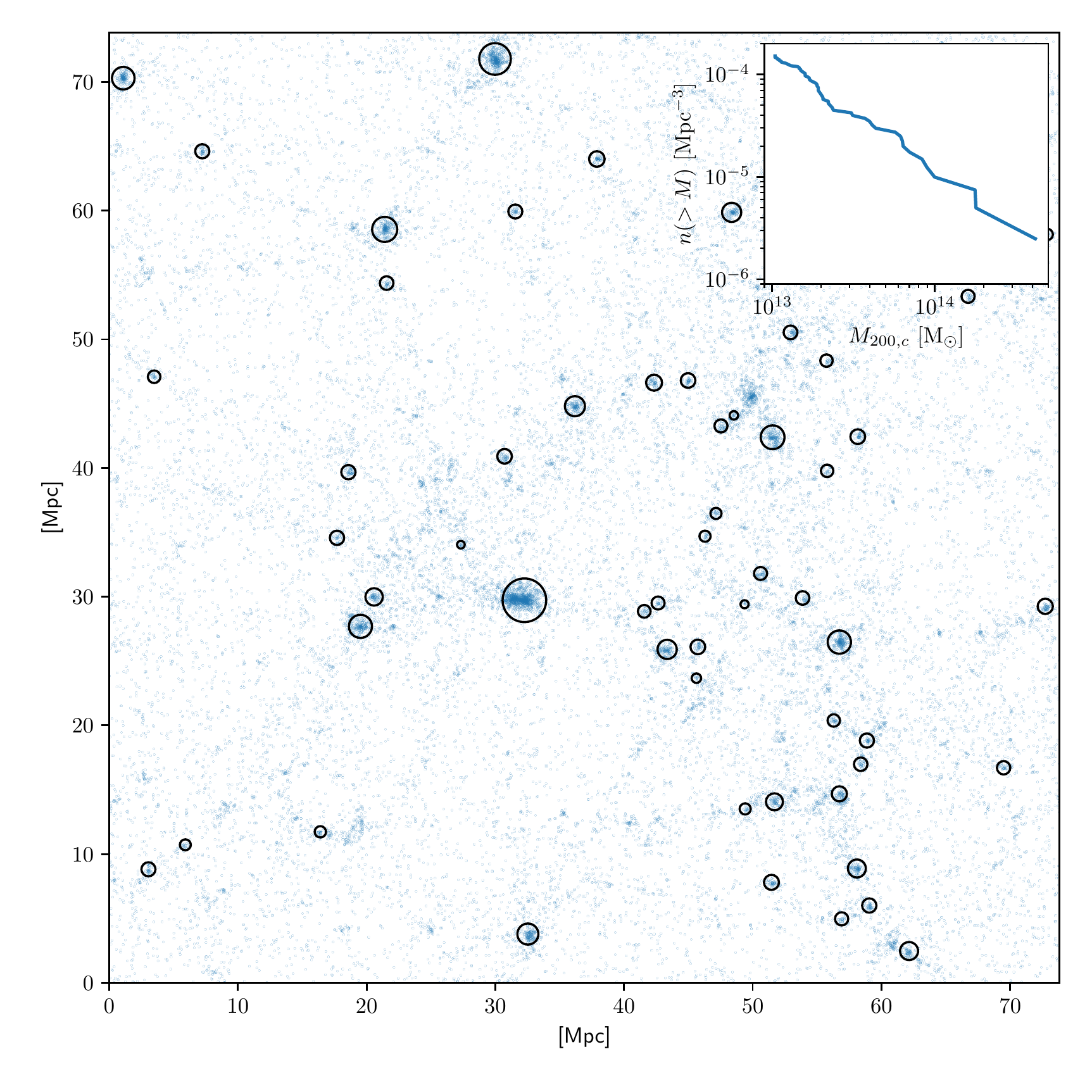}
  \caption{The cosmic large-scale structure in a gravity-only
    cosmological simulation. The blue dots denote the simulation
    particles of a very low-resolution ($32^3$) simulation, the black
    circles denote the halos identified by the \textsc{Subfind}
    algorithm, with sizes indicating their corresponding virial radii
    $R_{200,c}$. The inlay shows the halo mass function in the
    simulation volume.}
  \label{fig:cosmo_box_gravity_only_3d}
\end{figure}

One of the most important types of simulations for cosmology is
gravity-only cosmological volume simulations, one example being the
Millennium simulation \citep{Springel+2005}. In our test case of this
type of simulation, we employ a computational box of $50 h^{-1}$ Mpc
on a side and $32^3$ simulation particles to serve as a
computationally cheap example. This allows for a rapid calculation of the
solution, which is then compared to a previously obtained one. In
particular, the halo masses of the emerging structures are compared to
a reference run that used the same initial conditions. The resulting 
structure at redshift $z=0$ is shown in Figure~\ref{fig:cosmo_box_gravity_only_3d} 
with an inlay of the halo mass function. We note that
the box is too small and the statistical power too limited to yield a
meaningful comparison with fitting formulae for the halo mass
function. We ensured, however, that the test is actually sensitive to
the gravity calculation by varying the gravitational constant by one
percent. The resulting halo masses are then beyond the tolerance limit
we impose on the test. The creation script for this test also easily
allows for the creation of user-defined initial conditions via the
\textsc{N-GenIC} and \textsc{MUSIC} \citep{Hahn+2011} codes.

\subsection{Cosmological volume with gas cooling and star formation}

\begin{figure}
  \includegraphics{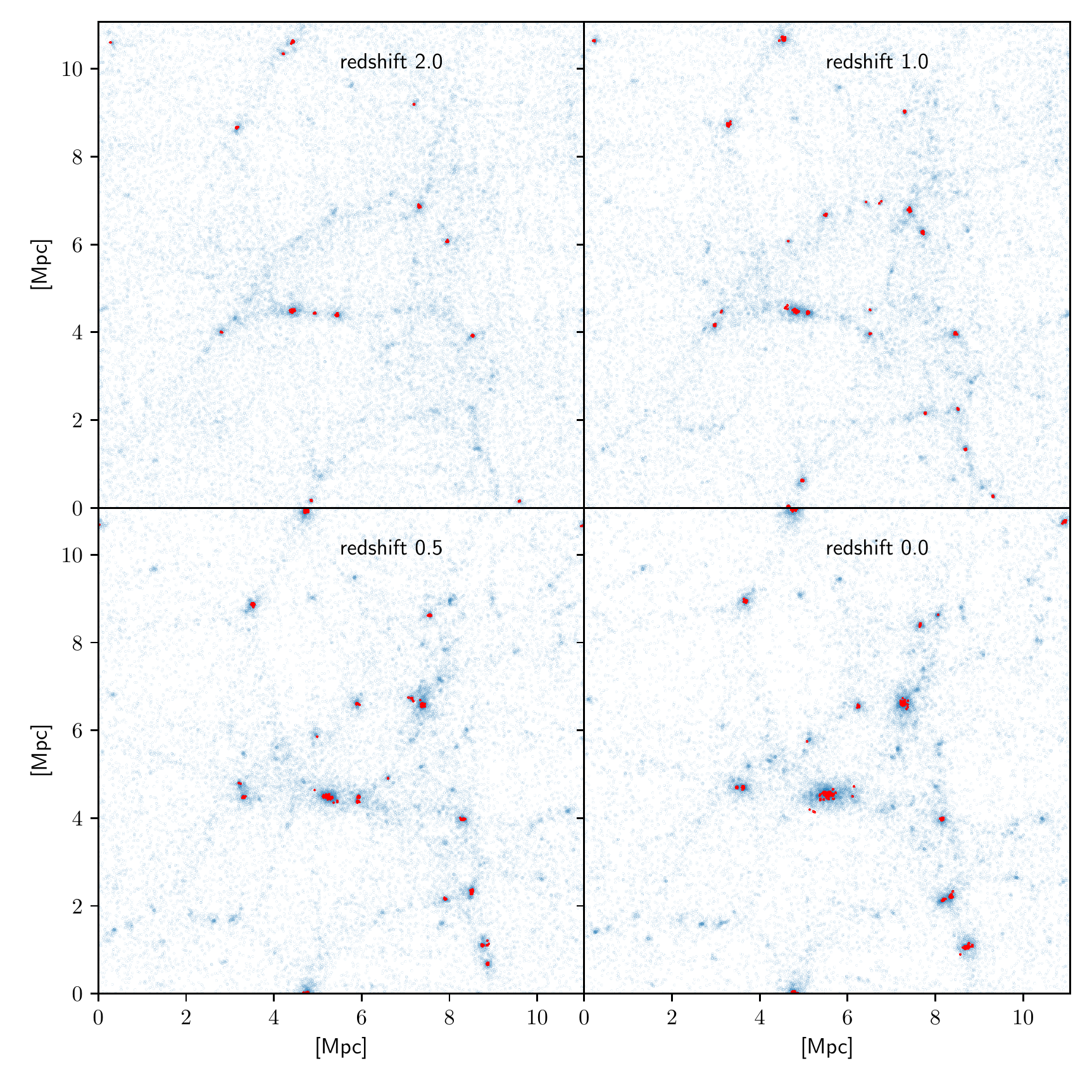}
  \caption{Cosmic dark matter (blue) and stellar (red) distributions
    at different redshifts for a cosmological volume simulation
    including gas cooling and star formation.}
   \label{fig:cosmo_box_star_formation_3d}
\end{figure}

Besides calculating gravitational interactions in cosmological volume
simulations, it is also possible with \textsc{Arepo} to include gas in
the initial conditions, and to allow for radiative cooling and the
formation of stars in gas exceeding a density threshold.  To resolve
this, one needs a significantly higher mass resolution than is used in the
previous test; thus, we employ a box of $7.5\, h^{-1}{\rm Mpc}$ on a
side but still use only $32^3$ collisionless particles.

\textsc{Arepo} reads these initial conditions and creates from them an
additional gas component according to the specified cosmic baryon and
matter density fractions. For this purpose, each dark matter particle
is split into a gas cell and a dark matter particle of reduced mass,
i.e., the simulation then has $2\times 32^3$ resolution elements
initially. Figure~\ref{fig:cosmo_box_star_formation_3d} shows that the 
stars (red) form at the centers of the dark matter (blue) overdensities.
The setup used here is very similar to one of the
cosmological simulations presented in \citet{Springel+2003}, and in
fact, the number of simulation resolution elements is comparable. One
of the key quantities we analyze is the star formation rate density as
a function of redshift, on which the verification is based. As before,
the solution is compared to a previous run, mainly because for this
quantity the correct solution is not known. Note that
the actual star formation rate in this particular test simulation will
exceed the rates deduced from observations substantially, due to the
absence of energetic feedback processes capable of driving galactic
outflows \citep[see][]{Springel+2003}. As in the previous example, the
creation script for this test also allows for the creation of
user-defined initial conditions via the \textsc{N-GenIC} and
\textsc{MUSIC} codes.

\subsection{Cosmological zoom simulation with gravity only}

\begin{figure}
  \includegraphics{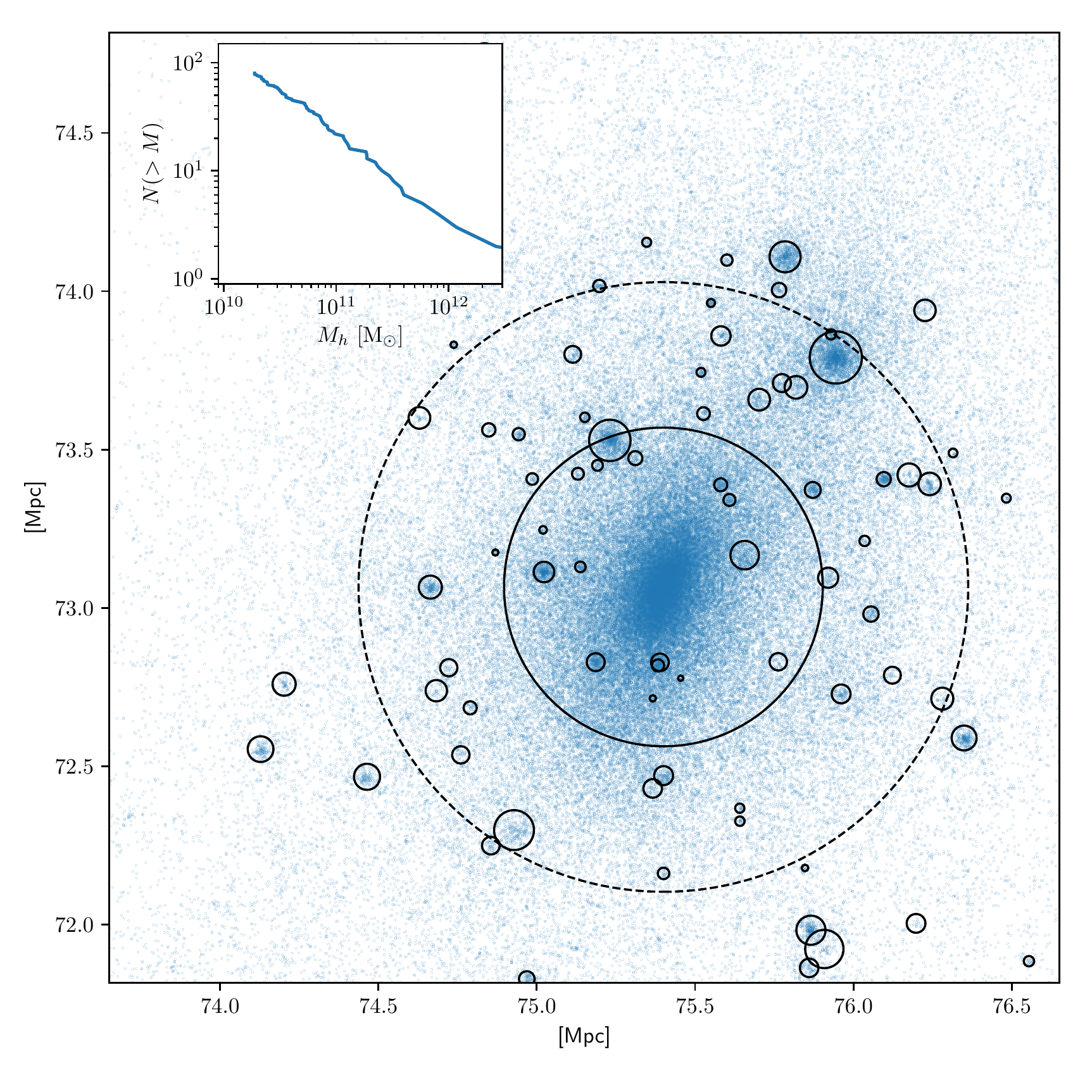}
  \caption{Halo structure of a zoom simulation of a $\sim 10^{14}$
    M$_\odot$ dark matter halo. The black solid circles denote the
    subhalos and their half mass radii as determined by the
    \textsc{Subfind} algorithm. The dashed circle denotes $R_{200,c}$
    of the main halo. The inlay shows the subhalo mass function within
    $2\, R_{200,c}$ of the main halo.}
  \label{fig:cosmo_zoom_gravity_only_3d}
\end{figure}

Another important type of simulation often used in galaxy formation
studies is cosmological `zoom' simulations. These are set up in a
similar way as the cosmological runs discussed earlier; however, they
employ a nonuniform mass resolution. In particular, for constructing
such zooms, a Lagrangian region for an object of interest is
identified as the region where all of the particles that end up in the
object at the final snapshot time originated from. This region is then
significantly refined in the initial conditions, whereas the rest of
the cosmological volume is sampled with poorer resolution. This allows
the large-scale tidal field originating from these regions to still be
captured without having to track its evolution in highly resolved
detail. In this way, it is possible to simulate individual objects in
their full cosmological context and afford significantly better mass
resolution for them.  \textsc{Arepo} has a number of features
specifically tuned for these kinds of simulations, in particular,
allowing the  high clustering of the computational workload to still
balance well.

We thus include a test of a zoom run to verify this functionality. 
Figure~\ref{fig:cosmo_zoom_gravity_only_3d} shows the redshift $z=0$
 particle distribution and the identified subhalos of the example
 halo, as well as the subhalo mass function in the inlay.
As a primary diagnostic of the test, we compare the subhalo masses
with a precomputed reference run with the same initial conditions as
the simulation itself.  The creation script in this case relies on the
\textsc{MUSIC} code.

\subsection{Isolated object, galaxy gravity only}

For studying the dynamics of an individual galaxy, a slightly
different kind of setup can be used. In particular, to study galactic
dynamics problems, one might be interested in setting up a stationary
collisionless galaxy and studying its evolution. Setting up such a galaxy
in equilibrium is a nontrivial task. Here we use the \textsc{GalIC}
code \citep{Yurin+2014} to create initial conditions of a compound
galaxy model, based on their model D, which has a spherically
symmetric dark matter halo and an exponential stellar disk.  Since
this is a stationary problem, it is straightforward to verify the
simulation by looking for the absence of secular trends. We here
measure the vertical scale height of the disk and its evolution in
time.

\subsection{Galaxy merger, including gas and star formation}

\begin{figure}
  \includegraphics{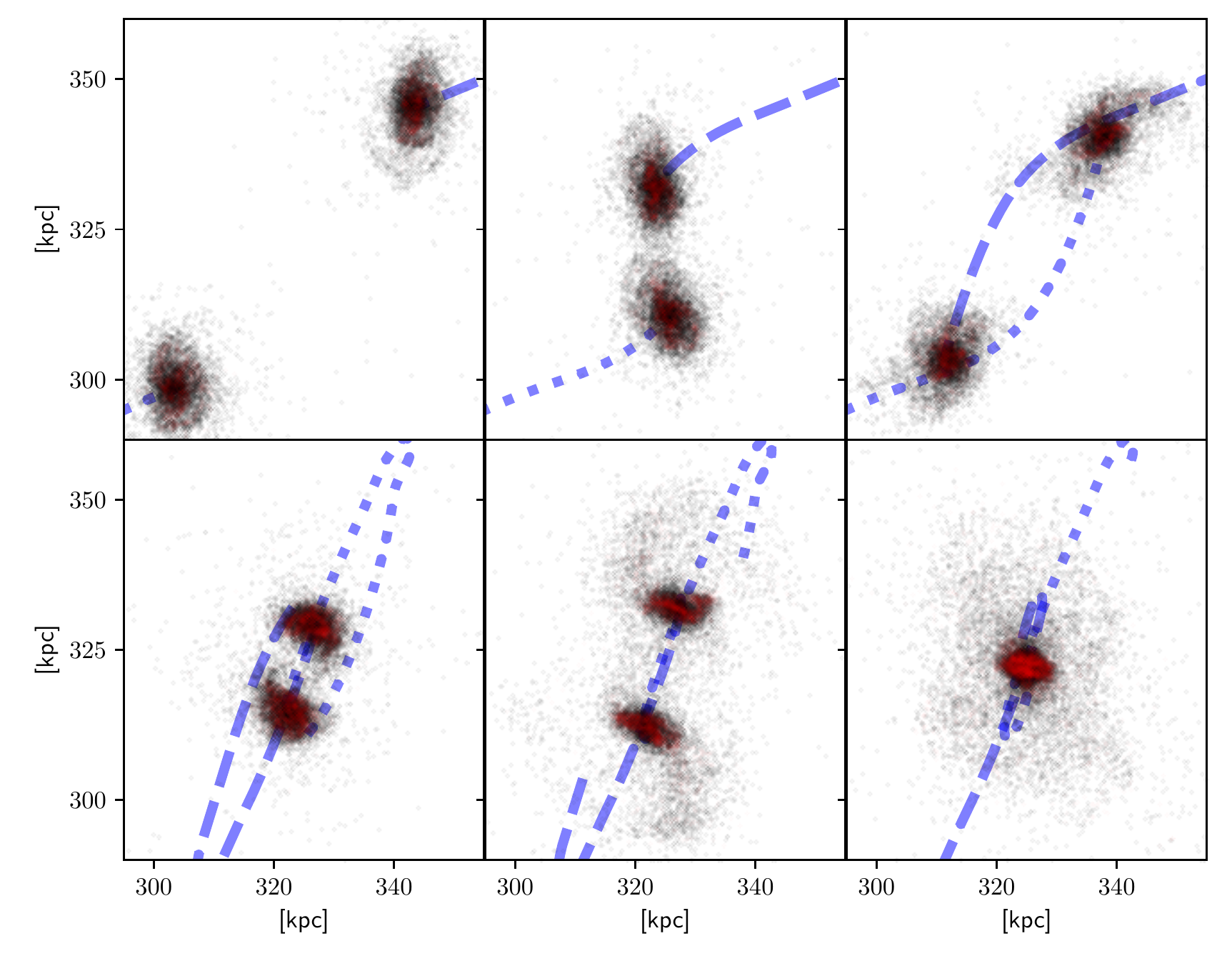}
  \caption{Time series of the position of existing (black) and newly formed (red) stars in a galaxy merger of two equal-mass galaxies. The blue lines indicate the trajectory of the center of mass of each of the galaxies. The upper row shows the first pericenter passage, and the lower row shows the second pericenter passage and the final merger.}
  \label{fig:galaxy_merger_star_formation_3d}
\end{figure}

A very informative numerical experiment to study galaxy
transformations consists of two isolated galaxies that are put on a
collision course.  Here we follow the setup used by
\citet{Springel+1999} and \citet{Springel+2005b} to study the merger of two disk
galaxies on a parabolic encounter orbit. Each galaxy includes a dark
matter halo, pre-existing stars, and gas. We note that the initial
conditions we use were designed originally for the SPH codes
\textsc{Gadget} and \textsc{Gadget}-2 \citep{Springel+2001,
  Springel2005}, which implies a number of small but important
differences in how the gas in the initial conditions is represented.

The most important difference is the difficulty of grid-based codes to
deal with a complete vacuum. While it is straightforwardly possible in
SPH codes to simply not place any gas particles into the background, a
grid code typically requires a nonvanishing gas density in these
regions, because otherwise, pesky stability issues arise due to tiny
floating point values and associated round-off errors.  As was
discussed earlier, we therefore use the special
\texttt{ADDBACKGROUNDGRID} mode of \textsc{Arepo} to convert the SPH
initial conditions to ones where the density field is defined
everywhere on a grid.  These are then evolved by \textsc{Arepo}
subsequently. Figure~\ref{fig:galaxy_merger_star_formation_3d} shows a time
series of the star particles in this simulation with the trajectories of the galaxies
indicated by the blue dotted and dashed lines. For verification, the resulting 
star formation rates are compared against
a reference solution.

\subsection{General remarks on initial condition generation}

The most important characteristic feature of \textsc{Arepo} compared
to Cartesian mesh codes or smoothed particle hydrodynamics codes is
that the choice of grid coordinates can be quite arbitrary and may
evolve in time in a smooth fashion. This allows a flexible adjustment
of the geometry of resolution elements and their local resolution to
the problem at hand.

As an example, in approximately spherically symmetric configurations,
such as in 3D hydrodynamic simulation of a star, the mesh can be set
up in a spherically symmetric fashion with radially adaptive cell
volumes, keeping the mass per cell approximately equal
\citep{Ohlmann+2017}. This flexibility in positioning the mesh
can, in principle, allow a simulation to have fewer cells at equal
resolutions than are possible with Cartesian adaptive mesh refinement techniques,
which are more limited in their mesh geometries. Yet, there are a
number of considerations that need to be observed to arrive at
high-quality initial conditions.

\textsc{Arepo} reads in positions of mesh-generating points and
creates the gas cells from them using a Voronoi tessellation for these
points. The mesh creation algorithm is described in detail in
\citet{Springel2010} and relies on a number of geometric predicates
that test whether points lie inside or outside of circum-circles around
tetrahedra (or triangles in 2D). The code here relies on an
unambiguous and correct answer to these geometric tests. This poses
additional challenges due to numerical floating point round-off when
degenerate mesh configurations are encountered, for example, when more
than four points lie exactly on the same sphere.  One case where this
occurs is a Cartesian configuration of the mesh-generating points.

While \textsc{Arepo} is able to cope with such situations, it needs to
employ exact arithmetic to robustly cope with the degeneracies in such
situations, slowing down the mesh construction. In practice, this
implies that starting out with a perfect Cartesian mesh leads to a
considerable amount of time spent in the first couple of mesh
constructions, until the mesh-generating points have started to slowly
dissolve the Cartesian configuration due to the fluid motion.  To
avoid this, one may use slight displacements from an ideal Cartesian
mesh, or better yet, use a honeycomb-like configuration. Note that it
is not necessary to supply the mass of each cell in the initial
conditions; one can instead also supply the density of each cell. This
greatly simplifies the placement of initial mesh-generating points, as
one does not need to know the resulting cell volumes of the Voronoi
tessellation ahead of time. Instead, one can simply supply the desired
primitive variables for each cell in the initial conditions, and the
code then computes the conserved quantities after the first mesh has
been constructed and the cell volumes are known.

For convergence tests or setups that require high-quality initial 
conditions, an additional complication has to be considered: in a 
finite-volume-based scheme, the primitive variables represent 
average values of the cells. Using the value of a desired input at 
the center of mass is usually a good enough approximation for each cell; 
however, for a Voronoi cell, the position of the mesh-generating point 
(which is specified in the initial conditions as the position) is in 
general \textit{not} the center of mass. Thus, using the primitive variables
at the mesh-generating point can degrade the quality of the initial 
conditions, depending on the used cell shapes.

\section{Code development aspects}
\label{sec:codedevelopment}

The \textsc{Arepo} code as presented here has in total almost $10^5$
lines of code, making it about a factor of five larger than the
\textsc{Gadget-2} code \citep{Springel2005}. The main contribution to
the increased complexity comes from the added algorithms for mesh
construction and hydrodynamics, but also other numerical parts have
become more sophisticated.  However, we stress that the code still
only gives basic functionality for astrophysical simulations, despite
being already of quite large size. In many research applications,
additional features such as, e.g., more sophisticated radiative
cooling routines, a tracking of different chemical elements, nuclear
reaction networks, treatments of radiative transfer, or more
sophisticated sub-grid models are highly desirable or
essential. Including such modules can easily increase the size of the
code by another factor of several. The corresponding work exceeds the
capacity of individual researchers, but becomes possible by a
community effort. We therefore would like to actively encourage
contributions by external developers in the form of code modules and
extensions developed independently from the original authors. In the
following, we discuss how we imagine such contributions can be made
based on the publicly released version of the code, and how this
relates to other versions of the code.

\subsection{The \textsc{Arepo} development version}

The original \textsc{Arepo} code grew considerably over the years both
in scope and number of developers/users. Currently, more than $150$
scientists have been given access to the developer version (for
comparison, in October 2014 this number was $50$). The source code
exceeds $3\times 10^5$ lines (less than $2\times 10^5$ lines in
October 2014). This growth in size, and perhaps more importantly in
the number of developers and users, has required us in recent
years to adopt a more formalized development workflow than the
anarchic model followed by \textsc{Gadget} and \textsc{Arepo} in the
past.

The development version of \textsc{Arepo} is now organized around a
master branch, in which all working, well-tested modules are
included. For the development of a specific feature, a dedicated
branch is opened. This branch is ideally kept in sync with the master
branch, and merged back to master once the feature is ready for
production level use. For the integration within the master branch,
the approval of one of the senior developers is required. For major
simulation projects, it has also proven practical to create a separate
branch with a stable code version dedicated to this project, and which
is then only updated with bug fixes.

Having all production level features included in the master branch
allows the main developers to make changes to foundations of the code
basis, such as, e.g., the MPI communication layer, and still ensure
functionality of all aspects of the code afterwards. However, one of
the disadvantages of this workflow is that the main code is prone to
become very crowded and complex eventually, and guidelines with
respect to code style and modularity need to be enforced as
well. Since the level of expertise in scientific code development
among the users and developers varies vastly, with most people in
computational astrophysics not having much experience with large
software development projects, this complexity can become a severe
obstacle for new developers to get started. It also requires
substantial amounts of time (proportional to the number of developers)
from senior developers to maintain the code. We therefore concluded
that this full-fledged development model is not well suited for the public
version of \textsc{Arepo}.

\subsection{Development philosophy of the public version}

Instead, we decided to provide a stable base version of the code, as
presented in this paper, which will be made available through a
powerful version control system (\textsc{git} in this case), but is
not intended for significant further development apart from bug fixes
and possibly the addition of further examples. Developments of new
features and extensions should be made in separate branches, forks, or
user's own repositories instead.  We further encourage developers to
provide access to their branch and/or their model as a patch to the
(stable) master branch, such that interested users can apply or
include it when needed. In this way, each model development is
independent of each other, and does need involvement of the original
\textsc{Arepo} authors. However, we encourage developers of new models
to contact us when they want to release their patches publicly, so
that we can maintain a list of available modules. The authors are also
happy to host stable versions as model branches next to the main
repository. It is obviously important in this approach that model
developments are based on a proper and consistent use of version
control software tools, so that the simple creation of patches is
possible, and so that updates of the master branch, in case they
happen, can easily be merged into them, too.

\subsection{Recommendations for additional models}

We note that the \textsc{Arepo} code still shares many resemblances in
its internal code structure with the \textsc{Gadget-2} and
\textsc{Gadget-3} codes. This will allow everyone who is familiar with
either of these codes to have an easy start with \textsc{Arepo}, and
makes it relatively straightforward to port modules written for any
version of \textsc{Gadget} to \textsc{Arepo}.

When turning such a port into a module patch for the public version,
it is highly advisable to consider a number of aspects. While the base
of the code is planned to be relatively static, there might be a large
number of different modules from other users, and some simulations
might require multiple modules.  It is therefore highly advisable to
design a module in such a way that the changes to the original code
are reduced to a small number of unavoidable function-calls, whereas
all of the extra functionality is implemented in new source code
files. This way of coding reduces the number of potential conflicts
between different modules significantly, and will, in many cases, allow
their joint use with no or little extra work.  Additionally, we
encourage developers to include (small) model tests of their own as an
example of the added functionality, and to allow easy verification of
the new module when other changes are applied as well. While it is not
possible to guarantee correct functionality in all cases, the
consistent use of such examples is by experience extremely helpful for
detecting usage problems, bugs, or model incompatibilities early on.

\subsection{Bug reporting and user support}

The continuing trend toward more complex astrophysical simulation
software makes training new users on how to use and improve the
software ever more important.  The main effort in making this public
release possible was in fact providing a complete documentation of the
source code and of various details of its usage, as well as preparing
small, out-of-the-box working examples as a guideline to set up new
simulations.

However, even the most complete user guide and the most extensive set
of examples will not cover all possible issues that arise. In the
past, remaining questions could be clarified by contacting the author
directly, e.g.~via email. This is, in principle, still possible, but it
has significant disadvantages.  The solution to a specific problem
arising from such an interaction is only available to the person that
contacted the author, and may not be archived/accessible in a
systematic way so that many people may run into the same problem time
and again. Also, answering all requests may require a significant
amount of time, which eventually decreases the quality of the support that
a given number of people can provide.

To improve the level of support, we host a support forum on the code's
website,\footnote{www.arepo-code.org/forums} 
where users can report their problems, and, importantly, 
also answer to problems of other users. This way, we hope to create a
supportive user community, in which more experienced users provide
help to less experienced ones. If successful, this will allow the user
base to grow without compromising the level of support that a new user can
get, as well as producing a public knowledge base for questions
related to the code's use. Bug reporting from users will also happen
via this forum, while the issue-tracking system of the code will serve
as listing confirmed bugs. Users will be notified about code updates
via a blog on the code's website.

\section{Summary and Conclusions}

We introduced the public open-source version of the \textsc{Arepo}
code and its underlying algorithms. \textsc{Arepo} calculates the
time evolution of initial-value problems of MHD using
a finite-volume approach on a moving unstructured Voronoi mesh. Other
open-source codes that adopted a similar approach include
\textsc{Rich} \citep{Yalinewich+2015} and \textsc{Shadowfax}
\citep{Vandenbroucke+2016}. Additionally, \textsc{Arepo} calculates
gravitational forces a using a tree-particle-mesh technique, making it
possible to include multiple species of collisionless (i.e.~only
gravitationally interacting) particles.

Time integration is performed for each element with its individual
local time step criterion, allowing for efficient calculations also in
situations with a large dynamic range in time. Additional gas physics
such as radiative cooling and a very simple model for the unresolved
interstellar medium are included as well, primarily serving as
guidelines for how such extra physics can be coupled to the main
code. A built-in FOF group finder and the substructure identification
algorithm \textsc{Subfind}, which may be used on-the-fly or in
post-processing, as well as routines to convert smoothed particle
hydrodynamics initial conditions to the appropriate equivalent
\textsc{Arepo}-input are further parts of the code base and should
simplify a productive use of the code especially for simulations of
cosmic structure formation.

With this public release, we hope to provide the community with a
useful tool for future research in theoretical astrophysics, and with
a good basis for further code development based on the concept of a
moving mesh. The source code is completely documented, and examples of
different complexity are provided to facilitate getting started with
the code. We also provide a support forum that is hosted on the code's
website as a platform to ask questions about the code, and to report
potential bugs if present. Ideally, this will grow into an active and
supportive user community that enriches the work of young students and
senior computational astrophysicists alike. Scientific developers are
encouraged to extend the code base in their own repositories, and
provide their modules via patches or public branches to the
community. While this can, in principle, be done completely
independently of the authors, we are also happy to host well-tested
branches as separate module branches and list them on the code's
website. It remains to be seen how well the particular model envisaged
here works in practice.  We are of course open to make adjustments if
they are indicated.

There is certainly no shortage of ideas for the next development steps to
improve the performance and capabilities of \textsc{Arepo}, some of
which have already been started.  Higher-order methods, both in
MHD \citep{Schaal+2015, Guillet+2019} and gravity
(Springel et al. 2020, in preparation) are very interesting for improving accuracy
for a given computational expense.  Also, \textsc{Arepo} has started
to include further physical effects such as radiation
\citep{Petkova+2011, Jaura+2018, Kannan+2019}, cosmic rays 
\citep{Pakmor+2016b, Pfrommer+2017}, as well as nonideal 
hydrodynamics and plasma-physics
effects \citep[][Berlok et al. 2020 in preparation]{Kannan+2016, Marinacci+2018b}. Significant further
research and additional development needs to be done to improve the
accuracy and universal applicability of these modules.

Other challenges lie in the distributed-memory parallelization layer,
which needs to be improved such that scalability can be extended to
much larger machine sizes. This is especially difficult for very
aggressive zoom simulations with their highly nonuniform resolution
and extreme dynamic range in density and time scales.  Similarly,
special-purpose hardware such as graphics processing units (GPUs) may
need to be embraced on the basis of redesigned algorithms that better
map to their streaming processors.  Another, so far sometimes
overlooked challenge is to minimize the energy used for the
solution. Optimizing this may well become very important for the use
of future supercomputers.

Continuous improvement of numerical methods and codes is clearly a
prerequisite to ensure progress in computational astrophysics and to
allow the technical advances in computer performance to be turned into
answers to open questions in theoretical astrophysics.  We are
convinced that the public release of \textsc{Arepo}, joining the
general trend for open, well-documented scientific software and
following the example of other codes, is a step in the right
direction, helping to support reproducibility and scientific progress
with computational methods in the light of ever more complex
simulation software.


\acknowledgments

The authors would like to thank the full user base of \textsc{Arepo}
for their continued encouragement to realize a public release of the
code, and for their long-standing efforts in putting the code to great
scientific use{ as well as the anonymous referee for the 
excellent suggestions that greatly improved the quality of this manuscript}. 
R.W., V.S. and R.P. would like to thank for financial
support by the Priority Programme 1648 ``SPPEXA'' of the German
Science Foundation, and by the European Research Council through
ERC-StG grant EXAGAL-308037.

%

\vspace{5mm}

\bibliographystyle{aasjournal}



\end{document}